**Ancient DNA in motion: Studying past human mobility and interactions by integrating archaeogenetics and archaeology**

**Authors**

Hannah M. Moots[1,2]*, Dilek Koptekin[3]*, Matthew P. Williams[4]*, Agathe Reingruber[5], Helena Jankovic Malmström[6$], Naoise Mac Sweeney[7$], Mehmet Somel[8$]

1) Naturhistoriska Riksmuseet (Swedish Museum of Natural History), Stockholm, Sweden
2) Stockholm University, Stockholm, Sweden
3) Department of Computational Biology, University of Lausanne, Lausanne, Switzerland
4) Department of Biology, Walla Walla University, Washington, USA
5) Institut für prähistorische Archäologie, Freie Universität Berlin, Berlin, Germany
6) Department of Organismal Biology; Human Evolution, Uppsala University, Uppsala, Sweden
7) Institute of Classical Archaeology, University of Vienna, Vienna, Austria
8) Department of Biological Sciences, Middle East Technical University, Ankara, Turkey

*: co-first authors
[$]: co-senior authors

Author for correspondence:
hannah.moots@su.se, naoise.macsweeney@univie.ac.at, msomel@metu.edu.tr

**Abstract**

Here, we present an overview of how archaeogenomic data can be used to investigate past mobilities and how it can be integrated with archaeological research to reconstruct models of human mobility at different scales. We first seek to explain the importance of a clear terminological and theoretical approach to mobility, discussing terms such as “migration”, “admixture”, “ancestry”, and “replacement”. We then describe the theoretical principles and state-of-the-art tools for inferring past mobility from ancient DNA data, including exploratory approaches and formal modeling, and the relevance of using diverse lines of evidence including allele-frequency patterns, haplotype-sharing, genetic relatedness, and uniparental markers. We further discuss the power and limitations of these methods for inferring different modes of mobility, such as large-scale and long-distance mobility, which we refer to as “migration”, and also small-scale and short-distance movements, which we argue constituted the bulk of past human mobility and may have had the major share in shaping cultural landscapes. We also describe inference of sex-biased mobility events. Finally, we present examples of joint analyses of genomic data with archaeological, bioarchaeological and historical evidence, demonstrating the potential of interdisciplinary analyses in building comprehensive models of past mobilities. We finish by calling for further integration across fields.

**Significance statement**

Ancient genomic data has been extensively used to construct models of past human mobility for more than a decade. Although early attempts were focused on migration-type events, there is now a growing appetite towards understanding human mobility at different scales and also with respect to its outcomes. This requires analyses of the archaeogenomic data at different levels, knowledge of the strengths and limits of analytical tools, and importantly, strong contextualisation of genetic data with archaeological and historical evidence. This article discusses these points and is intended for geneticists entering archaeogenomics and archaeologists navigating aDNA research, while also seeking to engage experienced researchers regarding challenges in mapping genetic patterns to historical outcomes.

## 1. Introduction: Understanding mobility in the past

Since the first ancient human genome was sequenced in 2010 (Rasmussen et al. 2010), many studies have leveraged genetic variation patterns to infer temporal signatures of human mobility (Williams and Huber 2025; Högberg et al. 2025; Anthony 2023). While the relationship between gene flow (migration) and genetic diversity is well characterized from a population genetic perspective (Schraiber and Akey 2015), its practical implementation in ancient DNA (aDNA) workflows and the integration of genetic findings with archaeological or anthropological models of past human mobility remain underexplored.

Here we discuss current approaches and methods for inferring past mobility events at different scales using archaeogenomic data and the relevance of interdisciplinary analyses. We aim to be accessible to a wide audience, and provide a glossary of technical terms (in bold font, underlined, and italic) as Supplementary Information.

Interest in past human mobilities spans a wide disciplinary spectrum. For geneticists, understanding mobility dynamics (of humans and other species) is interesting because these shape genetic diversity patterns and also because movement in space alters natural selection landscapes. From a social sciences and humanities perspective, mobility has, at times, played a role in sociocultural changes, innovations, and transformations. Indeed, movement has been a fundamental part of the human experience throughout history (Aldred 2020).

It should come as no surprise, then, that the study of past human mobilities has a long history. It was a core part of archaeological research in the early twentieth century, with large-scale migration often invoked as an explanation for cultural change as reflected in ***<u>pottery chronologies</u>***, and populations were often essentialized as the carriers of fixed and bounded "cultures". Markers of different populations were sought not only in cultural differences, but also in the morphological features of ancient skeletons studied through biodistance analysis (Nikita 2020). Racial, ethnic, and cultural groups were conflated, assumed to be naturally-occurring and fixed units. In ***<u>culture-historical</u>*** approaches,

"Prehistory was seen as a kind of global chessboard, with the various cultures as pieces shifting from square to square. The task of the archaeologist was simply to plot the moves" [Renfrew quoted in (Lewis-Kraus 2019)].

Many nationalistic ideologies of the early 20th century (including Nazi ideology) co-opted these theories. Following the horrors of the Second World War, however, questions of ancient migration were viewed as tainted, felt to be antithetical to global civil rights and decolonisation movements (Jones 1997). As these developments coincided with the scientific turn in archaeology, the focus shifted from characterising the movement of people to identifying the movement of things, mapping the distributions of both raw materials and cultural elements, including archeological artifacts, languages, and social practices (e.g. burial rituals). While such studies initially relied on studies of ***material culture*** style, these were soon complemented by various archaeometric analyses, such as chemical analyses to identify the provenance of materials (e.g. clay, metals), and stable isotope analyses conducted on organic matter (e.g. bones, seeds). Yet while complex theories were developed to explain material interconnectivities in the abstract (e.g. (Van Dommelen and Knapp 2010; Peregrine 1996), explicit engagement with the theme of human mobility remained relatively limited into the early twenty-first century (Anthony 1990; Hakenbeck 2008), in part because of a residual political discomfort described earlier.

The explicit study of past human mobility received a revival in the 2010s, when archaeogenomics began to provide information directly interpretable as human movement (Skoglund et al. 2012). Published ancient human genomes have well-surpassed 15,000 (Mallick et al. 2024) and the combined data contains detailed information about past mobility, especially in regions where favourable climatic conditions and biases in sampling effort (e.g. in Europe) have led to large skeletal collections to be sequenced. For archaeologists, this new impetus, coupled with the globalisation turn, has prompted renewed interest in all forms of past human mobilities, including but not limited to studies of aDNA (Aldred 2020; Burmeister 2016; Daniels 2022; Cabana and Clark 2020; Fernández-Götz et al. 2022; Schachner 2012; Anthony 2023; Gokcumen and Frachetti 2020).

These developments have created much excitement in academia and across the broader public. However, the rapid growth and cross-disciplinary nature of the field has also brought about terminological ambiguities and misunderstandings across disciplines, as recently highlighted (Crellin and Harris 2020; Frieman and Hofmann 2019; Patterson et al. 2022). We therefore start our review by discussing problems regarding terminology.

## 2. Terminology issues: "ancestry", "migration" and "replacement"

A central aspect of scientific research is the creation of a common language and frameworks by using standardised concepts, to avoid identical facts being described as different concepts, and vice versa. Émile Durkheim's call on sociologists is universally valid: "*The first step of the sociologist must therefore be to define the things [they examine], so that one knows and knows exactly what the problem is […]*." [(Durkheim 1894), pg. 131]. As an interdisciplinary group of researchers with different native tongues, we also had to agree on terminology from the beginning of our work. For example, the German-English dictionary www.leo.org suggests five different German translations of "migration": *Wanderung, Abwanderung, Auswanderung, Wegzug, Fortzug*. In addition, "emigration" is also translated as *Aussiedlung* or *Übersiedlung*, each with a slightly different connotation. This demonstrates the difficulty in being precise in wording, if the words themselves have different meanings.

"**Ancestry**" is one such difficult term used both in common parlance ("one's ancestors") and as a population genetic term (Mathieson and Scally 2020), where it can refer to the paths through an individual's pedigree (the theoretical definition), or the genetic affinity of an individual to specific reference groups (in practical use): "an individual carrying 50% X ancestry" (**Figure 1**) (Coop 2022). In the latter sense, "ancestry" refers to average genomic profiles, created by complex demographic processes of population splits, drift, and admixture (like language dialects). As such, ancestries, "***ancestry components***" or "ancestry sources" are subjectively defined (Coop 2022). For instance, "late Holocene Andean ancestry" refers to variant frequencies common in that period and region (given available genetic data) and defined based on semi-subjective temporal and spatial boundaries. These terms are hence semi-fluid: new genomic data or changing definitions of borders could alter the average profiles. Outside academia, however, "ancestries" may be heard as race-like entities, genetically distinct and fixed populations. Despite these caveats and calls for its replacement (Coop 2022), the term "ancestry" (referring to average genetic profiles) has become strongly embedded in demographic history analyses (**Section 3**).

Within the archaeogenomics literature "**ancestry**" is sometimes used in conjunction with identifiers relating to subsistence modes [e.g. "Western European Hunter-Gatherer (WHG) ancestry"] or cultural horizons (e.g. "Corded Ware ancestry", referring people associated with a pottery decoration style from 3000-2500 BCE Europe). Employing cultural labels to describe genetic groups is often done out of practicality, but conflates two independent phenomena, and can promote essentialist fallacies within and outside academia. An accurate and effective solution is to use labels that depend on location and period, instead of cultural affiliation (e.g. "12,000-8,000 BP Western Europe") (Eisenmann et al. 2018). Publications that do resort to cultural terminology for genetic groups should clearly acknowledge this shortcoming.

"**Migration**" is often used in aDNA studies as an umbrella to describe all types of human mobility. However, this varies from its usage in everyday parlance: permanent shifts of residential location across long distances. Within the disciplines of history and archaeology, "migration" usually implies large-scale movements (both in distance and in numbers), occurring quickly (over the course of months or years, rather than multiple generations). Some have argued for a further nuancing of this definition, suggesting that the term should be used to describe mobility events across large distances, irrespective of number of generations involved (Reingruber 2018). Others have also called for careful use of terms such as "migration" and "colonisation" (Hofmann et al. 2024; Furholt 2021; Frieman and Hofmann 2019; Hakenbeck 2019; Hofmann et al. 2023), treating these as only two extreme forms of human mobility, as established in sociology (Burmeister 2000).

In archaeogenomics, having a term reserved for large-scale, long-distance events that involve permanent residence may be helpful because the scale of mobility events shapes the genetic signatures detectable. Here, we use "migration" solely in such contexts (**Section 4a**), while we define **"Mobility"** as a more neutral umbrella term for all modes of human movement, including temporary movements, short-distance movements, and small-scale events. Such "background mobility" can also be genetically detected (**Section 4b**). Given sufficient data, researchers can further describe mobility events referring to scale (numbers of individuals), distance, and number of generations (duration).

Mobility events beget interactions between local and arriving individuals. Genetic admixture is sometimes referred to as **"replacement"** or "partial replacement". This implies antagonistic interactions, such as competition or violent conflict, and consequent loss of the autochthonous/preceding population. Such events did happen, but interpretations should then be supported by independent evidence (Gnecchi-Ruscone et al. 2024), not simply admixture models. Using "partial replacement" to describe admixture is further questionable, as it implies reduced reproductive success of local populations, which may be unfounded. For example, during the Neolithic spread in Europe (10-6 kya), there is a shift 100% WHG-like profiles to roughly 10% WHG- and 90% Anatolia-like profiles; this does not necessarily indicate 90% of locals were lost, particularly if WHG populations were initially much smaller than those coming from Anatolia (LaPolice et al. 2025). Similar considerations apply to Neanderthal–modern human interactions, where Neanderthal groups may have been absorbed into much larger human populations, rather than being replaced (Nielsen et al. 2017). "Replacement" is also a loaded term in current political discourse, as in the "Great Replacement Theory", a white nationalist conspiracy (Obaidi et al. 2022). Defaulting to violence as the driver of past interactions may inadvertently serve anti-immigrant political agendas.

The use of "**admixture**" to describe ***gene flow*** between populations has been criticised for implying the prior existence of pure genetic groups (Kampourakis and Peterson 2023). We agree with the criticism, but unfortunately, this term (like "ancestry") appears highly entrenched in the literature and would require significant effort to replace. We therefore encourage practitioners to use "admixture" with due care and awareness of the potential (ab)use of their research.

In summary, we note that the semantics of these terms (mobility, migration, replacement, admixture, ancestry, etc.) are constantly and semi-independently evolving in diverse academic fields and in the public sphere. We can address this challenge by having regular interdisciplinary discussions (e.g., in annual society meetings) to clarify terminology and to minimise conflict. We also bear responsibility for communicating our research with an appropriate awareness of the political reverberations of its language.

## 3. Archaeogenomic methods for studying past mobility

The analyses of ancient genomes provide insights into past mobility in several ways: by identifying admixture between genetically distinct groups, by identifying individuals whose genetic profiles suggest distant origins, or by finding relatives buried apart. Methods to capture admixture and differentiation patterns can be exploratory or involve formal testing; genetic relatedness is also analysed by formal tests. These analyses mainly use information from ***single nucleotide polymorphisms (SNPs)*** (variants) detected in ancient genomes, while some methods also employ ***haplotype*** information (i.e. DNA segments inferred computationally).

The methods we will cover here overlap with standard population genetic tools used on modern-day genomic data to investigate demographic processes such as population splits, ***genetic drift***, and admixture (**Figure 1**). Archaeogenomic applications involve further modifications to deal with large-scale missing data that characterise ancient genomes. Missingness is caused by DNA molecules being degraded and lost over time [reviewed in (Stoneking and Krause 2011; Orlando et al. 2021)]. Hence, the majority of ancient genomic data produced yet is on the scale of 0.1x depth-of-***coverage***. This corresponds to 10% of the genome, or 100,000 SNPs on a 1 million SNP panel. Consequently, for most ancient genomes the ***genotype*** information is not ***diploid*** but of ***pseudohaploid*** format, i.e., at a single position only one ***allele*** of an individual can be measured, from either the mother or the father, randomly per variant (conversely, high-coverage genomic data and diploid genotypes are standard in modern-day genomics). Ancient DNA data also contains specific errors, which require additional processing (Stoneking and Krause 2011; Orlando et al. 2021).

In this review, rather than delving into technical details of aDNA data and analytical tools for demographic inference or relatedness in depth [for this, see (Peter 2016; Harris and DeGiorgio 2017; Williams and Huber 2025; Patterson et al. 2012; Gopalan et al. 2022; Haber et al. 2016; Racimo et al. 2020; Lefeuvre et al. 2026; Aktürk et al. 2024)], we will provide an overview of the “typical” analytical pipeline in aDNA research for the purpose of studying mobility, how these methods have been used, and their limitations.

### 3.a. Exploratory analyses: dimensionality reduction (clustering) techniques based on genome-wide allele frequencies

Exploratory methods based on genome-wide allele frequencies provide hypothesis-free first insight into diversity patterns, and about demographic history. Because genomic datasets contain millions of variants, statistical and graphical summaries are required to detect such patterns. Since ***recombination*** allows different regions of ***autosomes*** to evolve partly independently (**Figure 2A**), each SNP provides semi-independent information about the thousands of ancestors of an individual a dozen generations ago (Ralph and Coop 2013). This can be complemented by information from ***uni-parental markers***, which strictly trace the maternal or the paternal line across generations (**Figure 2A**).

#### Common methods

Exploratory methods involve summarising either genotype data matrices or genetic similarity matrices into a few dimensions, rendering any patterns visually recognisable. The **Principal Component Analysis (PCA)** approach reduces information across millions of SNPs to a few dimensions, a.k.a. principal components (PCs) (**Figure 3**), which represent the most common similarity patterns across the genome. These similarity patterns, captured in PCs, frequently reflect spatiotemporal proximity in natural species, including humans (Novembre et al. 2008). Researchers often visualize the first two PC dimensions, which explain the highest variance, even though inspecting additional PCs could reveal finer-scale patterns of similarity (de Gennaro et al. 2025). A limitation is that standard PCA methods cannot deal with missing data, while most ancient genomes are ***low coverage***. Researchers thus usually calculate a reference PC space from a chosen subset of present-day genomes, and then “project” the ancient genomic profiles onto this space (Rasmussen et al. 2010).

Another dimension reduction approach involves first building a pairwise distance (or similarity) matrix among genomes in a dataset, and then summarising this matrix using **multidimensional scaling (MDS)**, usually in two dimensions. The pairwise distance metric could be the mean mismatch rate across pairs of genomes (P0, or $f_2$) or the inverse of a similarity measure, e.g. 1-***outgroup($f_3$)***. MDS clustering can be performed using pseudohaploid ancient genomes with high missingness without projection [e.g. (Fu et al. 2016)]. MDS or **hierarchical clustering** can also be performed using p-values from methods such as qpWave (Moots et al. 2023) or using haplotype-sharing metrics (Martiniano et al. 2017).

**ADMIXTURE** (Alexander et al. 2009) is a commonly used algorithm that assigns each individual a fraction of ancestry from K hypothetical ancestral populations (which can also be considered a K-dimensional ancestry space) (**Figure 3**). It may be performed supervised (ancestral populations defined by the user) or unsupervised (the algorithm identifies the most divergent K genetic profiles as the ancestral populations). Users typically analyse their dataset using several K values (e.g. 2, 3, …, 20) and may search for an optimal K via several approaches, though results with different K values are frequently reported together. Given the higher number of dimensions explored, ADMIXTURE can capture diversity patterns missed by PCA/MDS. These tools are therefore frequently used alongside each other (Skoglund et al. 2012).

There exist many alternatives or extensions to these algorithms. **OHANA** (Cheng et al. 2022) uses an alternative optimization to ADMIXTURE (Alexander et al. 2009), while **DyStruct** (Joseph and Pe'er 2019) uses the archaeological age of individuals for clustering. Studying spatiotemporal maps of ancestry proportion estimates (e.g. from ADMIXTURE) or genetic similarities (e.g. $f_3$, or the $f_4$ statistic that we explain below) is yet another common approach, e.g. (Broushaki et al. 2016; Allentoft et al. 2024; Yang et al. 2020). These descriptions reveal how genetic similarity patterns vary in space and time and can provide insight into mobility dynamics.

**Interpretation with respect to mobility.**

Key functions of exploratory analyses include 1) obtaining an overview of similarity patterns in the data, 2) generating hypotheses and 3) defining clusters of genetically similar individuals for downstream analyses (e.g. qpAdm), 4) identifying ***genetic outliers***, 5) detecting possible technical issues with the data. We list a number of specific patterns and their interpretations in terms of mobility:

- **Individuals clustering** in PCA/MDS/ADMIXTURE space is consistent with having similar demographic histories.
- Contemporary genomes from the same region forming **different clusters** implies distinct histories, or ***"population structure"***. This could arise from recent gene flow (without sufficient time to admix) (**Figure 1B**) or geographic or social constraints (**Section 5**).
- **Temporal shifts** in PCA/MDS/ADMIXTURE space when comparing genomes from a single region (say "X") can indicate **admixture**, e.g. mobility into X from a source Y with a different genetic profile, followed by admixture. Alternative explanations are also possible:

- Genetic drift can cause temporal shifts in PCA/MDS/ADMIXTURE space, if the drifted population (or their relatives) are used in calculating the PC space or ADMIXTURE components (Lawson et al. 2018).
  - ***Cryptic population structure*** and sampling different subpopulations from region X at different time periods can falsely appear as temporal shifts.
- Genomes from a region **retaining their positions** in PCA/MDS/ADMIXTURE space over time implies no major admixture from a single genetically distinct source.
- Individuals from region X that fall outside the bulk of contemporary genomes from X in PCA/MDS/ADMIXTURE space are termed ***genetic outliers*** (**Figure 4A**). They may represent first- or later-generation incomers, may have relatives within the "local" group, and be of high or low status.

**Limitations and best practices.**

- **Low-coverage:**
  - Low-coverage genomes may be falsely attributed divergent ancestry components in ADMIXTURE or appear as outliers in PCA, because information per SNP is minute and noisy due to recombination and drift. A minimum coverage threshold can help avoid this, although there is no universal value. Robustness of assignment to low coverage can be explored using ***downsampling experiments***.
- **Confounding factors - technical:**
  - Variant calls may also be subject to technical biases, such as allele frequency differences between data generated by ***SNP capture*** versus ***shotgun sequencing*** (Davidson et al. 2023; Rohland et al. 2022; Margaryan et al. 2020). Such technical differences may appear as real differences in demographic history. Batch effects from data generation and processing (e.g. datasets processed by different labs combined directly) can also create artificial differences. The issue will be compounded when effect sizes are small (when groups compared are genetically similar). Solutions include using only one data type (Koptekin et al. 2025) or subsets of SNPs less affected by technical biases (Rohland et al. 2022; Davidson et al. 2023; Fournier et al. 2025).
  - ***Reference biases*** *(Günther and Nettelblad 2019)* or ***ascertainment biases*** (Clark et al. 2005) may also distort similarity patterns.
- **Equifinality:**
  - The same patterns in PCA/MDS/ADMIXTURE space can be caused by distinct admixture histories. For instance, in PC space, population C being positioned between A and B can be explained by admixture between the latter two, but also admixture between unsampled source populations (**Figure 3**).
  - Admixture and drift can both cause shifts in PCA/MDS/ADMIXTURE space, but can be distinguished using formal admixture tests.
- **Subjective decisions:**
  - The choice of the reference set of genomes used to calculate the PC space impacts what patterns can be detected (Elhaik 2022). Likewise, in ADMIXTURE, different K values tell different stories. Various algorithms identify a single best K, but these may be biased towards choosing too small values (K=2) and may not be biologically meaningful (Do and Terhorst 2026; Coop 2021).
    - How clusters and outliers are defined shapes downstream analyses. However, defining clusters and outliers systematically can be challenging because many populations form clines instead of distinct groups in PCA/MDS/ADMIXTURE spaces. Statistical methods for identifying clusters and outliers vary in their performance depending on signal-to-noise ratios, and no standardised procedure yet exists.
- **Recommendations:**
    - Run multiple methods in parallel. Focus on patterns that are reproducible across multiple analyses for interpretation, and those supported formal statistical tests (see below).
    - Try alternative parameters (e.g. alternative reference sets in PCA, different K in ADMIXTURE) and report all results.
    - Interpret results with caution, considering that inferences are based on several assumptions, some difficult to evaluate.
    - Acknowledge these limitations and the subjective nature of the process in publications. This may be obvious to the population geneticist, but not necessarily so for a wider interdisciplinary audience.

### 3.b. Formal tests and models of admixture

Several methods explicitly model admixture events using allele-frequency correlations (Pickrell and Pritchard 2012; Patterson et al. 2012; Hofmanová et al. 2016). The most widely used set of tools are based on ***f-statistics*** and allow ***significance testing***, calculated based on the consistency of f-statistics across independent segments of the human genome (the “block jackknife”) (Patterson et al. 2012). Because human chromosomes are mosaics of segments inherited from different sets of ancestors, consistent admixture signals across the genome indicates a reproducible pattern. Moreover, f-statistics are generally robust to drift and therefore more readily interpretable as admixture (Patterson et al. 2012), as opposed to shifts in PCA/MDS/ADMIXTURE space. Groups used in these tests can represent a single individual’s genome or those of a group, often defined in exploratory analysis.

The **$f_4$-statistic** (and its close relative, the D-statistic) measures differences in genome-wide allele frequency correlation between pairs of individuals/groups from four populations (e.g. A, B, C, and O) (Patterson et al. 2012; Peter 2016) (**Figure 4B**). The $f_4$ is the correlation of A-C versus the correlation of B-C, given an ***outgroup*** O (which serves as a reference). The null hypothesis is that C is equally close to A and B, thus $f_4$ is 0. A positive and significant $f_4$ value (e.g. $p<0.01$) indicates asymmetry towards A, compatible with gene-flow from the relatives of C to the ancestors of A, though alternative scenarios are possible (Patterson et al. 2012; Peter 2016).

**qpWave** is an $f_4$-based method with several applications, and its most common use is comparing the genetic profiles of two groups (e.g. A and B), relative to reference populations ($R_1$, $R_2$, $R_3$, etc.) to test if A and B are ancestrally equivalent (Patterson et al. 2012), i.e. if the reference profiles $R_i$ are equally distant to A and B. None of the tests being rejected (e.g. $p>0.01$) can be interpreted as A and B having similar profiles.

**qpAdm** is another $f_4$-based tool for estimating whether and how two or more possible ancestral sources could have mixed to create a target population (Haak et al. 2015; Harney et al. 2021), e.g. if the genetic profile of target "T" may be explained by admixture between the sources $S_1$ and $S_2$. The procedure involves estimating the best combination of sources: e.g. 30% $S_1$ and 70% $S_2$. It further evaluates the fit of this estimated model profile "M" to the empirical profile T. qpAdm performs both estimation and evaluation based on $f_4$ statistics using $S_1$, $S_2$, T, M, and reference groups ($R_1$, $R_2$, $R_3$, etc.). The final tested hypothesis is that the M and T are equivalent relative to the reference set. A non-significant final p-value indicates that the model cannot be rejected, and may be useful. qpAdm has become the workhorse of admixture modeling over the last decade.

Other methods also exist for formal admixture testing. Rare Allele Sharing (**RAS)** (Huang et al. 2026) and **Twigstats** (Speidel et al. 2025) leverage rare and/or younger alleles to infer recent demographic events. **qpGraph** and **TreeMix** create phylogenies given individual/population genetic profiles and can incorporate admixture events across branches (Patterson et al. 2012; Pickrell and Pritchard 2012). Additional approaches involve reconstructing complex demographic histories with **maximum likelihood**, estimating gene flow, population size and split times simultaneously (Marchi et al. 2022). This depends on calculating the **site frequency spectrum** (SFS) from high-quality genomes (e.g. >10x coverage to allow diploid genotypes to be relatively reliably called), which limits its applicability. Alternatively, Approximate Bayesian Computation (ABC) can be used to compare among alternative demographic models using pseudohaploid SFS data and other summary statistics (Clemente et al. 2021; Tsoupas et al. 2025).

**Interpretation with respect to mobility.**

- Individuals or groups **indistinguishable in qpWave** can be treated as having similar ancestries. Similarity suggested by exploratory methods and supported by qpWave is likely reproducible.
- The $f_4$ test, qpWave, and qpAdm can all be used to infer admixture processes.
    - If contemporary genomes from the same region can be **distinguished in qpWave** and have **different qpAdm models**, this could indicate recent or past mobility into the region.
    - **Temporal changes in qpAdm ancestry proportions** within a region (assuming the region is genetically homogeneous) could represent mobility and admixture over time.
    - Individual genomes differing significantly from those of the same region and period could be considered "**outliers**".
    - $f_4$ tests and qpAdm modeling could further indicate the possible **sources of mobility**, such as Medieval-period genetic outliers in the Levant being connected with Europe (potential Crusaders) (Haber et al. 2019), or Native American admixture into Rapa Nui being connected with present-day populations of the Andes (Ioannidis et al. 2020; Moreno-Mayar et al. 2024).
    - Using qpGraph models, researchers can also infer admixture events that involve so-called "**ghost populations**" that have not yet been directly sampled. Such inferred groups have been reported for pre-Hispanic Mexico (Villa-Islas et al. 2023), for Early Holocene Southwest Asia ("Basal Eurasian ancestry") (Lazaridis et al. 2016), and for Tibet (T. Wang et al. 2025).

**Limitations and best practices.**

- **Confounding factors:**
    - f-statistic-based methods rely on allele frequency correlations, so technical biases (capture versus shotgun produced data, differences in data processing) can create artificial but statistically-significant signals.
    - f-statistic-based methods are theoretically expected to be robust to drift. Meanwhile, complex demographic scenarios with multiple admixture events can lead f-statistics to behave in unintuitive ways: e.g. two populations with similar demographic histories and ancestry profiles may show higher "affinity" to a much distant population than to each other (e.g. North African populations choosing Sardinia in $f_4$ tests over other North African populations) (Atağ et al. 2024).
    - Outgroup populations having unknown admixture histories with the modeled populations can cause false admixture signals (Flegontov et al. 2023).
    - f-statistics are generally considered robust to SNP ascertainment bias (Patterson et al. 2012), but not fully. Studying African populations with SNPs determined largely in non-African populations can generate false affinity signals (Bergström et al. 2020; Flegontov et al. 2023).
- **Subjective decisions:**
    - **Definitions of groups** (clusters, meta-populations) used in modeling will affect the inferred admixture patterns.
    - **Source and reference population choices** remain partly subjective (educated guesses, plus trial-and-error). Testing multiple plausible source/reference configurations ensures robustness.
    - Reference selection is also inherently subjective; having too few and/or inadequate references could elevate false positives, having too many could boost false negatives (Harney et al. 2021; Flegontova et al. 2025). One should try alternatives in parallel.
- **Model structure and limitations:**
    - Well-fitting qpAdm models do not necessarily represent the best fitting model (unless tested via tournament-type approaches, see below), let alone an actual admixture event.
    - qpAdm modeling is often performed using **distal sources**, i.e., explaining a population as combinations of groups from much earlier periods (e.g. modeling modern-day Europe as a mixture of genomes from W Europe, SW Asia, and E Europe of 10,000 years ago). Here, models are used in an exploratory fashion (Lazaridis et al. 2022). qpAdm may also be performed using **proximal sources**, possible source populations temporarily close to the target individual/population. These are more readily interpretable as plausible admixture events.
    - If the model sources (proxies) are genetically close to the real sources that created the target individual/group, the proportions of admixture will roughly represent the reproductive contribution of the admixing groups. Conversely, if one or more of model sources are distant from the real sources, the admixture proportions of the latter will be distributed among the proxy

proportions in sometimes unpredictable ways (Williams et al. 2024). A model of 50% source A and 50% B may also be constructed with admixture proportions 20-80% when the model includes different populations related to A and B (**Figure 3**).
    - Admixture events inferred by these models could be single pulse or multiple-generation events. qpAdm cannot distinguish between these alternatives (Harney et al. 2021; Williams et al. 2024; Flegontova et al. 2025).
- **Statistical interpretation and choosing models:**
    - The p-values calculated in $f_4$-tests, qpWave, and qpAdm are useful to decide about model validity. But in reality, their interpretation within a standard hypothesis testing framework is often not possible. Multiple dependent tests are often performed without multiple testing correction, and thus the chosen significance threshold (e.g. $|Z|>3$, or $p<0.001$) does not actually correspond to a Type I error probability. False negative rates in qpAdm (where models are proposed as acceptable based on non-rejection) are likewise unclear.
    - The p-value is a function of the signal intensity and of the amount of data (sample size). Two genomes with different profiles may be distinguishable by qpWave at high coverage but not at low coverage. A simple solution is to apply minimum coverage thresholds. Conversely, given sufficient coverage, the relative rank of p-values between models of a single analysis can be informative about their relative degree towards representing the admixture (Williams et al. 2024).
    - Another useful heuristic is the "tournament" approach (Narasimhan et al. 2019; Flegontova et al. 2025; Lazaridis et al. 2025) to compare two models (e.g. X+A and X+B) by using the alternative sources in the reference population list. For instance, if one uses B or A in the reference list for the models X+A or X+B, respectively, rejection of one model would be strong evidence for the alternative.

In summary, as with PCA/MDS/ADMIXTURE, the power to detect admixture in f-statistic-based tests depends on the amount of genetic differentiation between populations, the amount of data, and various noise sources. Admixture between genetically similar populations will be largely invisible, and conversely, confounding factors can create false signals of similarity or difference. For these reasons, f-statistics-based tools may *also* be treated largely as explanatory. Statistically supported models are working hypotheses, not proofs of events.

### 3.c. Further methods using autosomal allele-frequencies: modeling connections in time and space

#### Admixture timing

Dating of admixture events can be particularly useful for inferring timing of mobility. Estimates of mobility timing can help interpret changes in the material culture record and/or evaluate historical evidence. With dense temporal sampling within a region, the appearance of a new ancestry could indicate mobility/admixture times. Alternatively, several methods can statistically estimate admixture times, given admixed genomes and candidate sources (e.g. source populations supported by qpAdm), exploiting the fact that recombination breaks distinct chromosome segments every generation. The most popular tool is **DATES** (Chintalapati et al. 2022). It was, for instance, used to infer pre-European contact between Native Americans and Polynesians in Rapa Nui (Ioannidis et al. 2020; Moreno-Mayar et al. 2024). However, the algorithm is designed to model relatively simple admixture events (e.g. events with non-admixed sources) and may lead to biased estimates when assumptions are violated.

**Modeling mobility and admixture over space**

Several statistical models use geographic information for studying admixture patterns. **MOBEST** is a probabilistic method that identifies genetic outliers by constructing expected distributions of genetic variation in time and space, and comparing individual genomes buried at a specific time and location with these expected values (Schmid and Schiffels 2023). It can be used to predict potential sources of outlier individuals without using ancestry categories. **EEMS/FEEMS** uses allele frequency distributions across contemporary genomes in combination with their geographic locations to infer regions where gene flow rates were high or low in the past (Petkova et al. 2016; Marcus et al. 2021). The **$S_{max}$ statistic** (Loog et al. 2017) uses temporal and geographic data to estimate the degree of mobility within a region, based on how much observed genetic distances among groups can be explained by their temporal versus spatial distances. In addition, studying genetic similarity statistics over geography can reveal correlations with longitude or latitude, or ***ancestry clines*** over space. Such patterns can then be evaluated in comparison with spatially explicit population genetic simulations using methods such as **SLiM**, **Slendr** or **SPLATCHE3** (Petr et al. 2023; Currat et al. 2019; Haller et al. 2026).

**Temporal increases in diversity within and divergence between groups**

Growing diversity within groups over time can be an indication of gene flow from genetically distant groups into the population over time (Skoglund et al. 2014). This can be particularly informative when admixture sources are diverse and do not create visible shifts in ancestry components. Increasing levels of pairwise distance within populations, measured as 1 - outgroup($f_3$), over time, has thus been interpreted as a signal of such invisible gene flow (Koptekin et al. 2023). Meanwhile, increasing divergence between ***gene pools*** of two regions over time could indicate that one or both receive gene flow from genetically distinct sources, and decreasing divergence can indicate their homogenisation. Several studies have used **$F_{ST}$** as a measure of between-group divergence to study such temporal patterns, reporting lower $F_{ST}$ over time (Skoglund et al. 2014; Lazaridis et al. 2016; Feldman et al. 2021). Alternatively, 1 - outgroup($f_3$) can also be used as a divergence measure and may be more straightforward to interpret in relation to mobility than $F_{ST}$, because $f_3$ is not impacted by drift (Koptekin et al. 2023).

**Close genetic relatedness across distances**

Genetic relatives separated in space is a direct indication of mobility, of themselves and/or their direct ancestors. This signal can reveal mobility even when the individuals/groups involved are genetically homogeneous and therefore admixture not visible as ancestry changes. Asymmetric frequencies of relatives identified in a burial group, e.g. between females and males, or between social classes, can also indicate biased forms of mobility (e.g. matrilocality or patrilocality, or mobility involving only commoners). Estimating up to ***third-degree relatedness*** between pairs of individuals with limited genomic data (e.g. 0.05x coverage) is possible using a range of methods that employ different information sources and approaches, such as **READ/READv2** (Monroy Kuhn et al. 2018; Alaçamlı et al. 2024),

**KIN** (Popli et al. 2023), **ngsRelate** (Hanghøj et al. 2019), **BREADR** (Rohrlach et al. 2025), **GRUPS-rs** (Lefeuvre et al. 2024). Many use comparisons of mismatches between the pair and expected mismatches between unrelated pairs from that same population (or calculated from allele frequencies). First- and second-degree relatedness estimates with these methods can be generally reliable even with only 1000 SNPs, but tend to be more noisy when identifying third-degree relatedness, limited by the natural randomness in the recombination process (**Figure 2A**) (Aktürk et al. 2024; Lefeuvre et al. 2025). Relatedness estimation can be further complicated by inbreeding and complex reproductive patterns (Aktürk et al. 2024; Lefeuvre et al. 2025), or by lack of a homogeneous reference population. If the pair of individuals tested and the individuals used as the comparison sample do not share a homogeneous genetic background, if there is population structure, and/or if the profiles vary due to technical incompatibilities (e.g. shotgun versus capture), relatedness may be over- or under-estimated. To identify possible anomalies, researchers can subsample genomic data from the same individual and test if the subsets have a ***kinship coefficient*** ~0.5.

**3.d. IBD-sharing and distant relatedness between individuals.**

The methods described until now use DNA information in the form of SNP genotypes, allele frequencies, and allele frequency correlations between individuals/groups, with higher correlation indicating closer ancestry. DNA information can be alternatively studied in the form of chromosome chunks (haplotypes) inherited in identical state from a common ancestor, called ***identical-by-descent (IBD) segments*** **(Figure 2A)**. To infer such haplotypes and their sharing between pairs requires ***diploid*** and fully ***phased*** data. Raw aDNA data is naturally composed of short DNA fragments and contains large amounts of missing parts. Still, full aDNA segments can be statistically inferred using ***imputation*** and ***phasing*** algorithms, leveraging information from modern-day genomes to fill in the gaps.

The classical tool for imputation is **BEAGLE** (Browning and Browning 2007), although **GLIMPSE2** is an imputation algorithm optimized for low-coverage genomes (Rubinacci et al. 2021). Imputation can be applied to ancient genomes with sufficient coverage (>0.3x is a common threshold for shotgun-sequenced genomes) and require large reference panels that represent possible haplotypes, usually constructed using high-quality modern-day genomic data (e.g. the 1000 Genomes Dataset) (Sousa da Mota et al. 2023). Algorithms such as **IBDseq** (Browning and Browning 2013) can be used to estimate sharing of identical segments (IBD-sharing) across individuals/groups (Allentoft et al. 2024). The method **ancIBD** (Ringbauer et al. 2024) performs the same task using a Hidden Markov Model to overcome possible phasing errors and is currently widely used, although it can only reliably detect segments of minimum size 8 cM.

IBD-sharing can capture not only close but also distant relatedness between pairs of individuals, going beyond ***third-degree relatedness***, which is the limit of allele frequency-based methods. For instance, sharing ~1% of the genome in IBD (~50 cM) between a pair of individuals is consistent with 6th-8th degree-relatedness (with a wide confidence range that also depends on demography). Such a pair could be contemporary individuals with a great grandparent who lived ~100 years ago (assuming a 30-year generation time), among other possible constellations. Distant relatives identified by IBD-sharing buried geographically apart indicate mobility of themselves and/or their ancestors, and can reveal movement within genetically homogeneous populations.

IBD-sharing further informs on shared demographic history. **ChromoPainter** and **fineSTRUCTURE** (Lawson et al. 2012) leverage this and can be used in parallel with other allele frequency-based clustering and ancestry inference methods (Allentoft et al. 2024; Martiniano et al. 2017). **RFmix** (Maples et al. 2013) and **FLARE** (Browning et al. 2023) paint distinct ancestry segments in imputed genomes; the size of segments can provide information about admixture timing (longer segments being recent). However, the sensitivity of these algorithms to imputation and phasing errors is insufficiently explored.

Despite the high power of IBD-sharing methods for detecting connections, these still have important limitations.

- Current understanding of IBD-sharing or its lack between populations is limited: under what conditions does absence of IBD sharing between genomic samples from two populations signal lack of mobility?
- IBD-sharing between two populations does not indicate direct movement between them. Both segments may be inherited indirectly, from ancestors at a third location.
- Imputation reliability depends on how well the reference panel represents the haplotype diversity of the population. Imputation quality of genomes from Africa is currently lower than genomes from other regions (Sousa da Mota et al. 2023), owing to the failure of present reference panels in capturing the high haplotype diversity in Africa. Imputation of temporally distant populations (e.g. Upper Paleolithic humans) may also be less accurate, assuming some haplotypes have been lost.
- How much the genotype and phasing errors impact downstream inferences of distant links and population similarity with methods such as ancIBD, IBDseq, ChromoPainter, is a topic actively explored (Zavala et al. 2025; Ausmees et al. 2022; Purnomo et al. 2025; Tretmanis et al. 2023; Escobar-Rodríguez and Veeramah 2024; Çubukcu and Kılınç 2024; Sousa da Mota et al. 2023).

Because DNA segments and allele frequencies carry partly different sets of information, and analyses using both types of data have unique limitations, it may be a good strategy to use both in parallel and seek consistent results, while watching out for new ways to analyse and interpret IBD-sharing data.

## 4. Modes of mobility: What archaeogenetics can and cannot tell us

The methods discussed above, together with many further approaches not detailed here, growing datasets, and deepening interdisciplinary collaborations are allowing us to differentiate between diverse modes of mobility, with respect to the size of the moving group, the distances covered, the number of generations, directionality, the social characteristics of the mobile individuals. Events may involve large-numbers and long-distances (Section 4a) or may be more local and/or involve a small number of individuals (“background mobility”) (Section 4b). Sex-biased, age-biased, and status-biased variations in mobility can also be potentially investigated (Section 4c). Here we discuss how different types of genetic evidence can be combined to study these diverse mobility forms, as well as the underlying assumptions and limitations in mobility inference from archaeogenomic data (Section 4d). We discuss how these genetic signals can be integrated with non-genetic evidence in Section 5.

**4.a. Large-scale mobility over long distances - "migration"**

**Definition.** For our purposes, we can define "long-distance" and "large-scale" events as movements that happen at scales that could frequently produce visible changes in ancestry components in human populations during the Holocene: a minimum of thousand kilometers and a minimum migration rate of a few percent. In many parts of the world and in many periods, populations separated by at least a thousand kilometers have tended to carry readily differentiable genome-wide profiles (***"isolation-by-distance"***). This very rough scale can be inferred from genetic differentiation in present-day West Eurasian data (Novembre et al. 2008), and genetic differentiation was even higher in the early Holocene (Lazaridis et al. 2016; Antonio et al. 2024). Further, mixing between groups at a rate of ≥5% is often detectable as ancestry component changes, e.g. using qpAdm (Koptekin et al. 2025). Even lower admixture may also be detectable when the admixing groups are highly differentiated and noise is low.

**Detecting migration with genomic data.** The genetics toolkit described in Section 3, using up to millions of genome-wide markers, can reliably capture temporal shifts in ancestry components, indicative of migration. Drift, like mobility, can also cause shifts in PCA/ADMIXTURE space, but f4-statistics or qpAdm models are robust to drift under most conditions. Hence, temporal shifts in qpAdm-inferred ancestry components can be readily interpreted as evidence of past mobility. Such shifts can be captured even with a single individual's genome. However, if there exists un-identified population structure, and different subpopulations are sampled from different time periods, this can cause artificial signals of ancestry component shift.

**The geographic origins of migration.** This source of movement may be genetically inferred, with different levels of confidence.

- Ancestry components (sources) in admixture models can inform on the geographic origins of movement. These may be investigated by statistical comparison between alternative qpAdm models (Narasimhan et al. 2019; Flegontova et al. 2025; Lazaridis et al. 2025) or using full demographic models with alternative sources and gene flow (Marchi et al. 2022; Clemente et al. 2021).
- The precision of such inference depends on the breadth of geographical sampling, genetic differentiation among sources, and data quality. For instance, in Rome, the sources of East Mediterranean-related gene flow during the Imperial period (Antonio et al. 2019) still remain unclear, due to the relatively homogeneous gene pools in Greece, Anatolia, and the Levant during the early 1st millennium CE, and limited sample sizes available for this period from these regions.
- Genetic relatives (close or distant) buried in different locations can also be informative about sources of mobility. Distant relatives (IBD-sharing) at two locations is not evidence of a direct path of mobility between those locations, but can still provide a range of possibilities. For example, population movements from the Pontic-Caspian Steppe across West and Central Eurasia during the Bronze Age are reflected in a wide network of IBD sharing patterns centered around the Steppe (Allentoft et al. 2024; Lazaridis et al. 2025).

**The duration of migration.** The number of generations involved in migration is another interesting characteristic, which may inform on the motivators and organisation of mobility.

- If dense temporal data is available, a steady rate of change in ancestry proportions over time could indicate continuous mobility and admixture. An example is early Holocene European ("hunter-gatherer") ancestry increasing over time in Neolithic European communities (Lipson et al. 2017). Conversely, a temporally narrow shift in ancestry proportions could indicate a brief event followed by widespread admixture. For example Central Anatolia receives Upper Mesopotamian admixture by 7000 BCE, just before the advent of full-scale farming; this ancestry appears at even levels across all individuals studied yet, but does not increase thereafter, suggesting a single bout of mixing (Koptekin et al. 2025).
- Dense data could also reveal the appearance of genetic outliers as movement starts or the mixing of individuals with different backgrounds in pedigrees, such as individuals with early Holocene European ancestry in Neolithic European communities (Gamba et al. 2014; Gelabert et al. 2025; Mattila et al. 2026). Movements in the historic periods, such as East Mediterranean migration into Italy (Antonio et al. 2019, 2024) or the movement of Slavic groups into the Balkans could also be followed using genetic outliers (Olalde et al. 2023).
- Strontium and oxygen isotopic data can further help determine temporality (Section 5).

**The direction of migration/mobility.** Migration events can be unidirectional or involve back-and-forth movements. The latter could be identified through dense temporal sampling or through population genetic modeling. For instance the Silk Road(s), along which people have travelled and traded along since at least 200 BCE, have facilitated movement across East Asia and the Mediterranean, connecting three continents. There is ample evidence for cultural practices moving along these networks, and recent aDNA work has provided a first glimpse into how sustained, multi-directional mobility shaped populations connected to these routes (Bandyopadhyay et al. 2025; Jeong et al. 2020; Zhang et al. 2021; Franklin 2024).

**"Patterned mobility".** The archaegeonomic record from the last millennia has revealed high frequencies of genetic outliers across Europe, attesting to long-distance migration. Yet this widespread mobility has not resulted in a collapse of population genetic structure in the continent, as expected from simulations (Antonio et al. 2024). A possible explanation may be "patterned mobility" (Moots et al. 2023), where the frequency and path of long-distance mobility is shaped by historical, cultural and geographic factors. These reinforce connections between particular subregions while maintaining differentiation between others. For instance, the spread of Steppe-related ancestry into Europe was not simply a function of geogrpahic distance, be instead was faciliated by relative flat over-land terrain in central Europe, but more limited in Iberia by the Pyrenees (Olalde et al. 2019). In Rome, shift in ancestry through time reflected shifting political and mercanitle networks (Antonio et al. 2019).

**Population sizes involved in admixture.** The scale of mobility events (the numbers of incoming compared to local individuals) is an important but difficult question, one that can sometimes be addressed more readily through genetic evidence than archaeological or historical. While ancestry shifts can suggest the magnitude of incomer contributions relative to local groups, ancestry proportions are not direct measures of population size:

- Ancestry proportion estimates directly depend on the source proxies used, e.g. in qpAdm (Williams et al. 2024). Increasing genetic distance between the proxy and the real source will lower the estimate.

- Ancestry shifts only reflect reproductive contribution, not census sizes of groups. For instance, if access to resources may be unequal between locals and incomers, this could create a gap between census numbers and inferred proportions (Antonio et al. 2024).
- The shifts represent the long-term impact on the gene pool. A 5% contribution continuing for 5 generations would create roughly a similar genetic effect as 25% in 1 generation, although the two alternative scenarios may entail distinct social dynamics.

Admixing population sizes can be estimated by studying ***runs of homozygosity*** (**ROH**), where high loads of relatively short ROH indicate small population size (Ceballos et al. 2018; Ringbauer et al. 2021)[(Ceballos et al. 2021) and/or studying IBD-sharing statistics [e.g. **hapNe** (Fournier et al. 2023), **TTNe** (Huang et al. 2025)]. A similar approach involves employing large numbers of summary statistics within an ABC framework: a recent example used this method to infer a 5-times population size difference between Neolithic-related populations and local hunter-gatherer-related populations in Europe (Tsoupas et al. 2025). Naturally, these methods depend on simplifying assumptions (e.g., no population structure). How violations of these assumptions may impact results may be explored with population genetic simulations.

**“Parallel communities”.** Large-scale migration may be followed by widespread admixture. Alternatively, the local and incoming communities may avoid admixture, which can create population structure and may be captured as two parallel genetic profiles. An interesting case is captured in cemeteries associated with Avar culture in medieval Hungary and Austria. Here, incoming groups with East Asian ancestry showed limited admixture with genetically local communities over centuries, despite local groups also having adopted Avar cultural practices (Gnecchi-Ruscone et al. 2024; K. Wang et al. 2025). These groups appear to have avoided ***endogamy*** by practicing female exogamy among communities with similar genetic ancestries (rather than their closest neighbours), sometimes 100s of kilometers away. More generally, without exogamy practices, constraining mating within a small community can create high levels of ROH. Examples include the Iron Age Carthaginian site of Kerkouane (in present-day Tunisia), where several individuals with ancestry from the East Mediterranean showed signs of high endogamy, likely linked to social restrictions on partner choice related to ideas of ‘Greek’ culture and identity (Moots et al. 2023; Ringbauer et al. 2025). In this case, cultural and genetic ancestry differences may have acted as parallel barriers to reproduction.

**Using uniparental markers to infer large-scale migration.** Large-scale migration can also be studied through changes in uniparental marker frequencies, given sufficiently large sample sizes; this may provide additional resolution if admixing groups are genetically similar. For instance, a shift is seen in Y-chromosome ***haplogroups*** in Late Neolithic Scandinavia, when I1 nearly replaces R1a, while autosomal ancestry proportions remain stable (Allentoft et al. 2024). However, caution is warranted here, as such shifts can also be driven by differences in reproductive success and drift (Guyon et al. 2024). Additionally mitochondrial haplogroups have also been using, in complement with ancient and modern autosomal genetics to demonstrate the continuity of Indigenous ancestries in Puerto Rico to the present day, which subverted the colonial narrative of "complete replacement" (Nieves-Colón et al. 2020).

**4.b. Small-scale and/or short-distance mobilities**
Although we lack direct estimates, it is intuitive to assume that most mobility in human societies involves individuals or small groups (e.g. <1% migration rate) moving over relatively small distances (<1000 km). This would likely be observed both in mobile foraging groups and in sedentary societies. Such events can be driven by economic opportunities, social expectations, seeking partners, etc., and can occur as one-off events or be regular and cyclical. It is possible that the predominant vector of cultural transmission in human societies is such small-scale and short-distance “background” mobility. Perhaps less frequently, large groups may also move across short distances (e.g. after natural disasters).

Within archaeology and history, these kinds of regional and local mobilities are emerging as a growing focus of research (Schachner 2012; Aldred 2020; Riva 2025). Although archaeogenomics has predominantly focused on migration events (Section 4a), smaller-scale and shorter-distance mobilities is now also a growing area of archaeogenomic research.
Small-scale and/or short-distance mobilities will not leave readily-detectable changes in ancestry components, but can still be captured in genetic data.

**Diversity changes.** Studying a regional population over time, imagine that ancestry proportions remain stable, but allele or haplotype diversity levels gradually increase. This could be explained by low-level mobility and admixture from distant sources happening simultaneously. This is observed e.g. in Anatolia between 3000 BCE to 1000 CE, where ancestry composition appears constant while diversity grows (Koptekin et al. 2023).

**Distantly buried relatives.** Genetic relatives buried geographically apart (i.e. at different sites or regions) can reveal low-level mobility, including mobility over short distances happening in a homogeneous genetic background. Examples include close (1st- to 3rd-degree) relatives buried >100 km apart in Late Neolithic, Iron Age, and Viking-period cemeteries, directly evidencing long-distance travel (Margaryan et al. 2020; da Silva et al. 2026; Gretzinger et al. 2024). These reports are primarily from Europe, that is likely a result of a majority of the published ancient DNA data at present being from Europe [as of 2022, over 65% of published genomes were from that region (Mallick et al. 2024)]. As ancient DNA datasets grow, these questions can be explored in other contexts.

**Genetic outliers.** Individuals with genetic profiles distinct from their contemporaries (without broader ancestry shifts in the population) are often interpreted as migrants or their descendants from genetically differentiated regions. Examples include the “Vittrup man” in Denmark, who carried hunter-gatherer-related ancestry linked to southern Scandinavia; isotope data further supported non-local origins and a shift from a marine-based to a farmer-like diet during his lifetime (Fischer et al. 2024). Another example is a Bronze Age Levantine woman with genetic ties to Central Asia, whose osteological profile suggests a life of servile labour (Skourtanioti et al. 2020).

**Pedigree outliers. I**ndividuals lacking close relatives, or connected only through descendants rather than ancestors, may represent incomers, potentially reflecting partner exchange. In Bronze Age Central Europe, high-status women identified as pedigree outliers likely exemplify such mobility (Mittnik et al. 2019). Similarly, at the Frälsegården megalithic tomb, a woman appears to have married into a large male-linked FBC pedigree and had children there, while her two brothers were buried in a nearby megalithic tomb (Seersholm et al. 2024).

**4.c. Demographic variation: sex and sex-biased mobility**

In many societies, the kinds of movement may differ by social status or by age, and unravelling such mobility forms requires the co-interpretation of archaeogenetic and other types of evidence (Section 5). Meanwhile, sex-biased mobilities may be directly visible from genetic data.

**Genetic sex inference.** Genetic sexing of ancient individuals can be achieved readily using DNA sequencing coverage data: XX individuals (biologically female) show comparable coverage on chrX and the autosomes, whereas XY individuals (often biologically male) show roughly half the autosomal coverage on chrX, and detectable coverage on ***chromosome Y*** (chrY) (**Figure 2**) (Lamnidis et al. 2018). Individuals with other sex chromosome karyotypes, including XXY, XXX, X0, or X0 mosaicism, have also been identified using archaeogenomic data (Rohrlach et al. 2024; Gresky 2024; Anastasiadou et al. 2024; Yüncü et al. 2025; Villalba-Mouco et al. 2021; Roca-Rada et al. 2022).

**Sex and gender in past societies.** Genetic sex information interpreted with archaeological context can illuminate how mobility, burial practices, diet, health, and other factors may have varied by sex. In contrast, studying gender - understood as a set of socially constituted and performed practices that may or may not align with prevailing norms (Butler 2002) - is much more difficult to study in past societies. Gender is expressed through culturally specific behaviors that may not straightforwardly map onto binary genetic categories. Because gender roles vary substantially among present modern-day societies and have also varied in the past, studying them in the past requires careful integration of archaeological context with archaeogenetic data (see **Section 5** below). Interpreting sex-biased signals as gendered practices requires additional consideration (Seferidou and Atağ 2026; Guyon et al. 2024).

**Inferring sex-biased mobility.** Sex-biased mobility patterns may arise as specific residence patterns, such as patrilocality or matrilocality, which describe movement related to marriage/reproduction, or other forms of sex-biased mobility and admixture. Residence patterns may be inferred from sex differences in the frequency of genetic relatives within communities or genetic links among generations in reconstructed pedigrees. An excess of adult male-male relatives and male-biased connections across generations have been interpreted as patrilocality in 5th to 2nd millennium BCE cemeteries in West/Central Eurasia (Rivollat et al. 2023; Mittnik et al. 2019; Chyleński et al. 2023; Blöcher et al. 2023; da Silva et al. 2026), in 2nd millennium BCE China (Chen et al. 2025), and among Avars in 6th-9th century CE Europe (Gnecchi-Ruscone et al. 2024). The opposite patterns have also been reported, suggesting matrilocal practices in communities in Neolithic Anatolia (Yüncü et al. 2025), Neolithic China (J. Wang et al. 2025), and in Iron Age Britain (Cassidy et al. 2025). Sex differences in the frequencies of genetic or pedigree outliers (described earlier) can be a parallel signal of sex-biased exogamy.

In addition, sex-biased mobility can be studied via diversity patterns. One approach compares changes on the ***X chromosome (chrX)*** and ***autosomes*** (**Figure 2B**): larger change on the autosomes versus chrX could signal male-biased admixture, such as male-biased admixture during the Pontic-Caspian Steppe-related Bronze Age migrations (Goldberg et al. 2017; Olalde et al. 2026; Clemente et al. 2021). Interpretation of such signals may be aided by comparisons with population genetic simulations (Goldberg et al. 2017), or with qualitative comparison of admixture events in time: e.g. if, over time, changes happen more on the autosomes relative to chrX, this could both indicate increasing male-bias or decreasing female-bias in admixture, as reported for Southwest Asia (Koptekin et al. 2023).

A parallel information source for studying sex-biased processes are the ***mitochondrial DNA*** (mtDNA) and the chrY (**Figure 2**). These are highly variable and non-recombining loci inherited uniparentally, often summarised as distinct ***haplogroups***. Sharing mtDNA or chrY haplogroups between distant groups can indicate maternal or paternal connections (i.e. female or male movement), respectively. If two populations converge over time in their mtDNA composition but not their chrY composition, this could indicate female-biased mobility, and vice versa. Male-biased East-Asian admixture in Mongolia within the last 1000 years has thus been inferred from patterns of chrY haplogroup sharing ((Bandyopadhyay et al. 2025; Jeong et al. 2020; Zhang et al. 2021; Franklin 2024)). At the cemetery or house level, high diversity in chrY versus mtDNA has been interpreted as matrilocality [e.g. (J. Wang et al. 2025)(Yüncü et al. 2025)], and the opposite, as patrilocality [e.g. (Rivollat et al. 2023; Szécsényi-Nagy et al. 2015)].

### 4.d. Limits to archaeogenetic inferences of mobility and equifinality issues

Inferences about mobility depend heavily on inferring ancestry component changes, inferring relatedness, and identifying outliers, all of which can be limited by a number of factors.

**Subjective choices.** Decisions about a) individuals to be included in a dataset for comparative analyses and, b) individuals to be grouped together as “populations” for admixture analyses shape downstream results (patterns emerging in PCA/ADMIXTURE and admixture model outcomes). These early choices therefore shape inferences about population continuity vs. admixture, sources and magnitude of admixture, presence of genetic outliers, etc. Decisions that may feel “data-driven” are inadvertently subjective to some degree and constrain possible outcomes. Performing exploratory investigation extensively, considering alternative groupings, and repeating modeling with alternative choices will increase robustness of conclusions. Researchers should also acknowledge uncertainty and avoid overstatements (such as “proving” or “disproving” historical models).

**Noise and bias in estimates.** Natural randomness in recombination and drift, and high degrees of missingness in ancient genomes can introduce noise in estimates, while technical factors can introduce biases [e.g. (Margaryan et al. 2020; Günther and Nettelblad 2019; Günther et al. 2025)] (**Sections 3.a and 3.b**). Inferring genetic relatedness through allele-frequency similarity or IBD-sharing via imputation are also subject to diverse sources of noise and bias [e.g. (Aktürk et al. 2024; Sousa da Mota et al. 2023; Ringbauer et al. 2024). Using multiple tools and alternative data processing approaches in parallel and focusing on reproducible findings will improve robustness.

**Population invisibility and structure.** Some populations (or population sectors) may be invisible to archaeology, e.g., due to taphonomic processes or research biases, or to archaeogenomics, e.g., due to burial treatments precluding aDNA extraction, such as cremation. If there exist genetic differences among sampled and unsampled subpopulations (cryptic population structure) and if different periods are unintentionally represented by different genetic populations, this can create false signatures of temporal ancestry shifts. The best practice here is that geneticists, archaeologists and anthropologists jointly consider various possibilities. Population substructure can also bias relatedness estimates.

**Further challenges in mobility inference.** There remain major challenges in characterizing mobility events using archaeogenomics data and the current analytical tools.

- Dating past admixture events and inferring their duration (number of generations) is one such challenge. Admixture dating tools model this process as a single bout event (Chintalapati et al. 2022). Inferring duration requires dense temporal sampling and jointly considering genetic results with archaeological/bioarchaeological information [e.g. (Fort and Pérez-Losada 2024)].
- Ancestry clines are observed in many regions [e.g. the north-south cline in East Asia (Yang et al. 2020) or the east-west cline described for Tibet (Wang et al. 2023), or both longitudinal and latitudinal differentiation in Europe (Novembre et al. 2008)]. These can arise through background mobility and admixture with neighbours over long durations (the standard isolation-by-distance model). But they can also be created by migration events involving increasing admixture with locals during the event's progression, as inferred for the Neolithic spread in Europe (Tsoupas et al. 2025). Likewise, complex migration patterns, such as back-and-forth migration, may leave signatures hard to disentangle. Increasing the density of spatiotemporal sampling and using multiple population genetics signatures to compare with spatially explicit simulations could help choose among alternative models.
- Inferring the proportions of incomers and locals, or the exact proportions of females and males is challenging (**Section 4.a**). Employing statistics calculated from multiple markers (**Figure 2**) and using allele-frequency- and haplotype-based statistics in combination and population genetic models of sex-biased admixture [e.g. (Musharoff et al. 2019)] could be useful strategies.
- The provenance(s) of movement and admixture is another difficult problem and requires dense geographical sampling, sufficient genetic differentiation, and limited population structure. Sharing short IBD segments between two regions can be suggestive but does not necessarily show direct mobility between them (as this could arise via admixture from a third shared source).
- Mobility and admixture events that involve genetically similar (often geographically proximal) sources will not be reflected as changes in ancestry. They can be detected by identifying close relatives, but this depends on large sample sizes and serendipity.
- Although uniparental marker diversity can be helpful in inferring sex-biased admixture patterns, data analysis and interpretation warrants caution. First, variable missingness in aDNA data can cause haplogroups to be identified at different levels (e.g. I2 versus I2a1). Second, beyond mobility/admixture, genetic drift can readily cause shifts in gene pool composition of uniparental markers, and the amount of drift is a function of reproductive success, which may differ between sexes (Guyon et al. 2025). For instance, low diversity in chrY in a cemetery may indicate patrilocality in that community but also a male-specific bottleneck in the past (the latter scenario can be investigated by comparing diversity patterns across a wider region).

### 5. Co-interpreting aDNA and cultural and bioarchaeological evidence

Integrating aDNA-based findings with insights and evidence from allied fields can allow us to develop fuller and more nuanced interpretations. Non-genetic evidence about past human mobility can come from a range of sources:

- **Material culture and cultural forms**, including objects, their physical form and styles, the ways they were made and deployed, the remains of architecture, monuments, and other human installations, as well as mortuary and other ritual practices. These sources can bear testimony to human movement either directly or indirectly. For example, in prehistory, the extraction and circulation of obsidian or metals required human agency, and therefore constitutes direct evidence for human mobility. Human mobility is also implied by the spread of cultural practices such as certain stylistic preferences (e.g., in the decoration of ceramics) or social practices (e.g., mortuary rituals). However, the mode, demographic scale, temporal scale, and geographical scale of these mobilities cannot automatically be assumed.
- **Written records**, either in the form of retrospective historiographical literary texts (e.g., histories, chronicles, epic poetry, etc.), or as primary historical documents (e.g., census documents, population registers, etc.). These may address human mobility directly (e.g., describing population movements, individual travels, or recording arrival/departure) or indirectly (e.g., graffiti attesting the arrival of a new language/script in an area, implying some form of human mobility and interaction in order to transfer knowledge of that language/script, although this kind of knowledge transfer can be complex and multi-stage, not necessarily requiring the long distance movement of individuals or groups).
- **The distribution of languages**, either present or past, can also inform on the history of mobility. Languages are an element of culture (see 'cultural forms' above) and can be transmitted (like other cultural forms) both vertically and horizontally across generations. As such, their dispersal involves some degree of movement. But languages can also spread with limited movement through horizontal transfer, which often involves bilingualism.
- **Landscape and environmental records,** including variation in pollen types preserved in sedimentary samples, shifts in soil chemistry between stratigraphic layers, and changing settlement patterns over time, can provide indirect evidence for human mobility. These patterns can point to moments of significant demographic change of occupation density and of agricultural strategies. The appearance of new plant and animal species in the environmental record may also be indicative of human mobility.
- **Human bioarchaeology**, through the study of skeletal remains, can offer direct evidence for human movement. The analysis of strontium ($^{87}Sr/^{86}Sr$) and oxygen ($^{18}O/^{16}O$ or $\delta^{18}O$) from bones and teeth can identify individuals who moved within their lifetimes. Sr isotope data has been particularly widely used in combination with genetic data to infer first versus later generation newcomers. For instance, several genetic outliers identified in Late Bronze Age Britain were also Sr outliers, with different C14 dates, suggesting continuous migration over multiple generations (Patterson et al. 2022). Bioarchaeology can likewise inform on lifetime experience and social status of moving individuals, and thus the drivers of mobility, such as being forced or being driven by economic opportunities.

- ***Metagenomic data***, i.e. information about microbe strains that lived in the human body can be another source of indirect evidence for mobility. Both pathogens and commensal microbes shared between interacting people provide evidence of connections (Eisenhofer et al. 2019).

All of these sources of data are complex and are the subjects of academic specialisation. Therefore, successful co-interpretation of aDNA evidence with these data sources requires interdisciplinary collaboration.

Perhaps the most common form of this interdisciplinary working is the study of mobility using material culture and genetic information at the individual level. Both genetic outliers and imported objects within a burial can be indicators of human mobility, the first constituting direct evidence, the second constituting indirect evidence. For archaeogeneticists accustomed to working with direct evidence for human mobility, it is important to bear in mind the nuances of dealing with this indirect evidence: the dead do not bury themselves, and the mortuary sphere is often a highly ritualised setting where social identities are constructed and negotiated. In the mortuary context, an object such as a bowl, strand of eggshell beads, or a blade, may not have been made, owned, or even used by the person they were buried with. Take for instance, an infant buried with a sword or any other object beyond their years to wield (Pearson 2000). Such an item would shed light more on the burying community than on the individual interred, and can potentially tell us about this community's ideas about status, kinship, gender roles, etc., which can be challenging to reconstruct.

One well-known example involves the Viking-age burial in grave Bj 581 in Birka, present-day Sweden. When first excavated in 1878, the burial was classified as an "elite warrior grave" and assumed to be biologically male. Indeed, Bj 581 was one of the wealthiest of more than 1100 graves excavated at the site, and one of only 2 burials that included the full complement of shields, spearheads, knives, arrowheads, axes, and swords. The high status nature of the individual was confirmed by other objects suggesting long-distance travel: a set of weights for, two horse skeletons and harness, and a set of fine clothes in the fashion of Eurasian Steppe riders. Assumptions about the individual were upturned, however, when archaeogeneticists found that this individual was genetically female (Hedenstierna-Jonson et al. 2017). This study helped shed light on the present-day gender assumptions that had been imposed upon this individual due to the inclusion of a sword in her burial. This work was combined with Sr isotope analysis that suggested the individual was not born locally, suggesting mobility. Archaeogenetic analyses have recently prompted re-readings of historical sources which mention female Viking warriors and travellers, which had frequently been dismissed as unrealistic (Price et al. 2019). This showcases how triangulation between material culture, text, bioarchaeology, and genetics can be transformative.

At another scale of analysis, the co-interpretation of archaeogenetic and other forms of evidence allows us to explore structures of kinship and community dynamics. In Neolithic Anatolia, for example, this has permitted a far improved understanding of possible household composition, by studying genetic relatedness among burials within the same buildings, suggesting the flexibility of kinship rules with respect to genetic links (Yaka et al. 2021; Yüncü et al. 2025). This work has also provided support for the long-disputed hypotheses about female-centered practices in early farming societies in Southwest Asia and specifically in Çatalhöyük, initially inferred from the predominance of female figurines (Mellaart 1965). Recent genetic analyses suggested that mobility among houses followed matrilocal-like patterns, i.e. male exogamy within the settlement (Yüncü et al. 2025). Related work includes the recent reinterpretation of archaeogenomic inferences of female exogamy under patrilocality in Europe, pointing out that female mobility among settlements may be considered not simply as a reproduction-related practice but one that involved women as knowledge transmitters, trade networks, independent travellers and females with political influence (Streiffert Eikeland 2025).

Further developments involve comparative analyses of material culture similarity and genetic similarity patterns, with the strict assumption that cultural histories and genetic histories are two independent historical processes that may sometimes overlap to different degrees. The bulk of such work involves qualitative comparisons that infer cultural transmission coupled with high or limited gene flow (i.e. background mobility). Examples of the latter type include the spread of Neolithic lifeways in West Anatolia (Koptekin et al. 2025), in the Baltic (Jones et al. 2017), and in North Africa (Simões et al. 2023), the spread of megalithic cultures in Europe (da Silva et al. 2026), or the adoption of Avar culture by local communities in Central Europe (Gnecchi-Ruscone et al. 2024).

Beyond qualitative comparisons, quantitative co-analyses of archaeological and archaeogenetic data are also emerging. For instance, researchers studied the Neolithic expansion in Europe by combining the rates of genetic admixture inferred from ancient genomes (10-15% admixture of local groups to incoming farmers by the end of the expansion) and radiocarbon dates of farming settlements across the continent, compared with large numbers of agent-based simulations (LaPolice et al. 2025). They estimated an upper limit of 2.5% of cultural adoption of Neolithic lifeways by local individuals, per generation.

Another team quantitatively compared genetic and material culture similarities (burial types, tools, architecture, etc.) across 16 settlements in the Neolithic East Mediterranean (Koptekin et al. 2025). Spatial proximity, but not genetic admixture, appeared as the main determinant of cultural similarities, suggesting cultural similarity being shaped by background mobility. While these analyses are inevitably reductionist and depend on a vast number of assumptions, the direction they are pointing towards is exciting, showing novel paths to contextualise genetic data with archaeology.

## 6. Conclusion: The way forward

The examples discussed provide a snapshot of a broader trend toward better integration of archaeogenomic data with archaeological evidence. Human aDNA studies are increasingly going beyond broad descriptions of admixture history of “ancestry” groups, and starting to explain aspects of history, including the causes and consequences of mobility patterns. In this article, we would like to propose three key future directions for the field.

1. The first is the current notable trend towards archaeogenomic studies focused on mobility of short distances and small scales (including so-called “kinship studies”), rather than relatively rare large-scale and long-distance events. This is exciting as such background mobility processes could be a major driver of cultural exchange among human communities, and a factor shaping social dynamics (see **Section 4b**). However, while genetic data can easily capture signals of migration-type events, it is less revealing about small-scale and short-distance mobility, which require high-quantity and high-quality data to detect. Natural noise in biological processes, technical factors, and missingness further limit or confound detection of small-scale events. Hence the need to use exploratory, modeling and simulation tools in combination and to focus on patterns supported by different lines of analyses (**Sections 3, 4**).

2. Another important direction is the consolidation, expansion, and accessibility of data sets. Increasing spatial and temporal sampling density is an obvious path to increasing accuracy of mobility estimates. But increasing sample sizes comes with the cost of destructive sampling. There is increasing awareness of minimizing the impacts of destructive sampling, and the importance of ethical practices for permit and community engagement. This also includes countering dominant practices of helicopter/parachute science in aDNA, where research is conducted without involvement of local archaeologists, bioarchaeologists, and geneticists. But we can also do much good simply by collating and making more widely accessible the existing data. Data integration requires careful documentation with common standards. Archaeogeneticists should publish detailed contextual information, including excavation IDs, specific bone elements used, numbers of excavated burials, samples with poor DNA preservation, and data ambiguities. This enables future studies to integrate genetic data with other evidence, e.g. isotopes or updated archaeological information. Initiatives such as Poseidon (Schmid et al. 2024) and MIxS-MInAS (https://www.mixs-minas.org/) have been established to record and track this information in a standardized way, and it would be for the whole community's interest to support and improve such community databases for compiling meta-information. Another key point is transparency about which bones are destructively sampled, enabling complementary analyses on remaining skeletal elements. Combining multiple lines of evidence from the same skeleton (e.g. aDNA, radiocarbon dating, isotopes for mobility and diet, osteology, pathogen and artefact analyses, and calculus) provides greater insight than sampling different individuals for separate analyses.

3. The final future direction is that of interdisciplinary working. To create meaningful models, archaeogeneticists and archaeologists need to use multiple lines of genetic evidence and combine this with other evidence types (from archaeology, linguistics, anthropology, and history) to co-interpret possible models of mobility. Such interdisciplinary analysis should by now be common practice, and performed with due attention to the sophistication and histories of research in the relevant fields. A failure to engage properly in interdisciplinary consultation can result in poor science, and can even be politically dangerous. Within the 15 year history of archaeogenomics, we have multiple examples where aDNA evidence has been used as material to support essentialist and racist attitudes. Triangulation depends on establishing and maintaining collaborative relationships. While this may be "slower science", it appears as the healthy route forward. We may even explore going beyond interdisciplinary collaboration and merging disciplines entirely. By integrating archaeogenetics within the long history of scholarship on human mobility, we can begin to use the same theoretical and conceptual frameworks across both archaeogenetic and other classes of data.

As we consider these directions for where we might be going in the future, it is perhaps also salient to reconsider how we got where we are now. With the processual turn in archaeology in the mid-20th century, it was argued that “pots don’t equal people”. This grew out of the recognition that the movement of material culture styles of objects did not necessarily entail large-scale migration or population replacement, but that there were many modes of mobility for objects, styles, and ideas. This led to the insight that culture is not essential, with a fixed “Culture” being the exclusive preserve of a bounded population group. Rather, culture changes continually over time, and the borders drawn between one cultural group and another are always subjective, fluid, and situationally dependent. Culture, we told ourselves, did not automatically map onto ethnicity. Yet in the late 20th century, it was recognised that ethnicity also was not fixed or primordial, but also situational, changeable, and socially constructed, with the distinctions drawn between ethnic groups being subjective. Ethnicity, we confirmed, did not automatically map onto anything as essentializing and physiologically based as “race”. In the last decades, it was further recognised that the boundaries drawn between racial groups were also fluid, situational, and socially constructed, and were not rooted in essential underlying biology. Race and racialised categories, we affirmed, were imaginary biology and did not map to genetics. Now, within aDNA studies, we can also acknowledge the fluidity, subjectivity and socially constructed nature of our genetic categorisations and methods. Genetic groups, no less than groups defined racially, ethnically, or culturally, are not natural - they are our own inventions.

As our understanding of humanity develops in ever tightening concentric circles, we understand better the various elements that go into the making of us - the pots, the practices, the genes. But we should not forget that while pots are not people, genes are not people either.

**Acknowledgements**

We thank Vagıf Mammedzada, Melih Kaplan, Ömür Berker Şenses, Muhammed Sıddık Kılıç for helpful suggestions. M.S. was supported by the VR Center of Excellence, and Center for the Human Past under the Swedish Research Council grant no. 2022-06620_VR and a Wenner-Gren Visiting Research Fellowship (GFOh2024-0045).

**Figures**

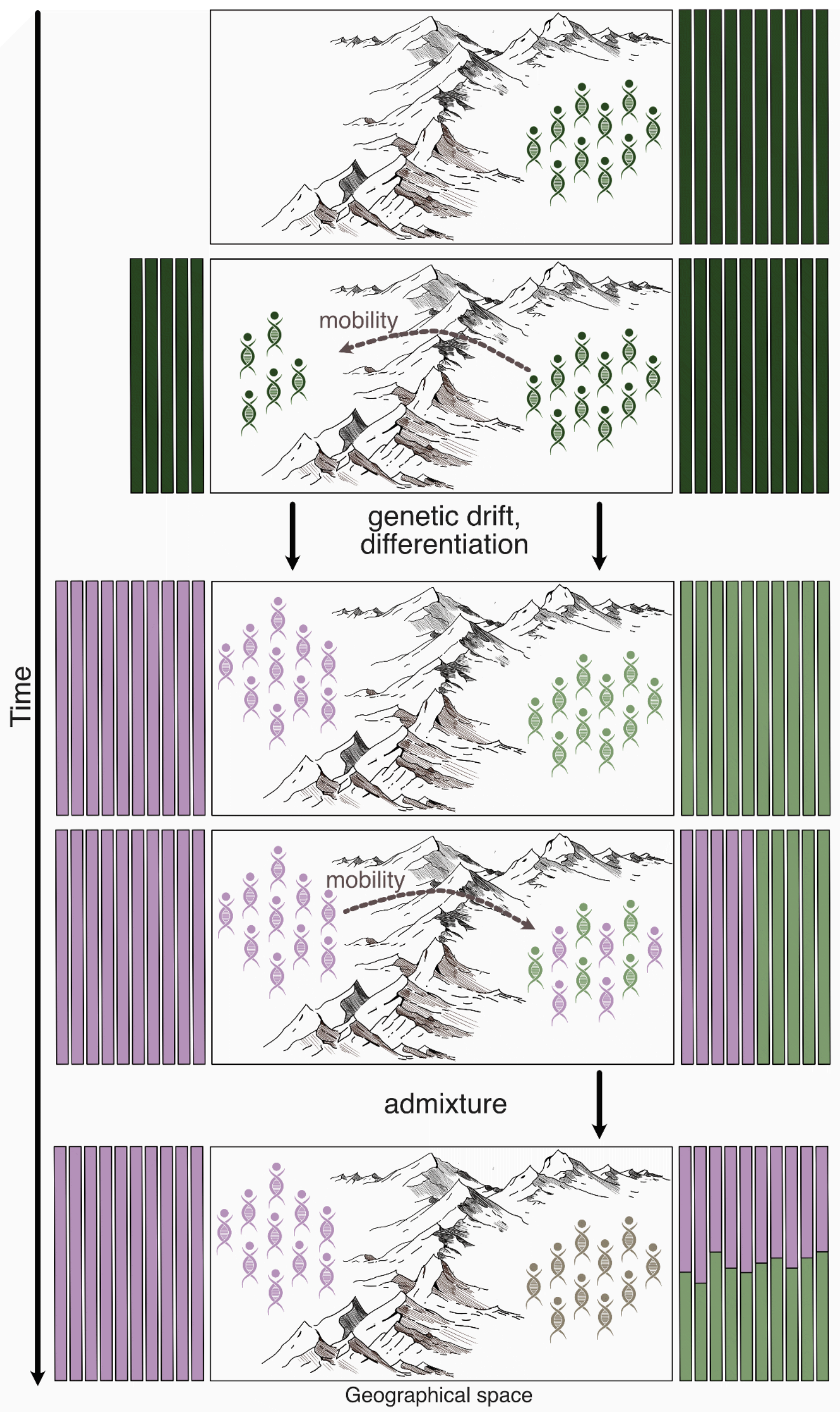

**Figure 1:** A schematic description of drift and admixture in time and geographical space. The coloured bars on the right and left of the main panels describe the admixture (or ancestry) profiles of the individuals on either side of the mountain range, of the type that could be obtained by the softwares ADMIXTURE or qpAdm. The appearance of new colours represent genetic differentiation through drift or unidirectional mobility followed by admixture.

**Alt text:** A figure showing genetic processes at 5 windows in time. In the first time-window, there is a single population of individuals (shown in dark green), all on the right side of a mountain range. In the second time window, some individuals from the initial population move across the mountain range, and there are two genetically similar populations (shown in dark green). In the third panel, these two populations have differentiated through genetic drift - they are now shown in light green and light purple. In the fourth panel, some individuals from the light purple population move and join in the light green population on the right side of the mountains. In the fifth panel, several generations have passed and now individuals on the right have admixtured. The individuals are colored brown and an admixture plot on the right side of the plot shows that their ancestry can be modelled and 50% light green and 50% light purple.

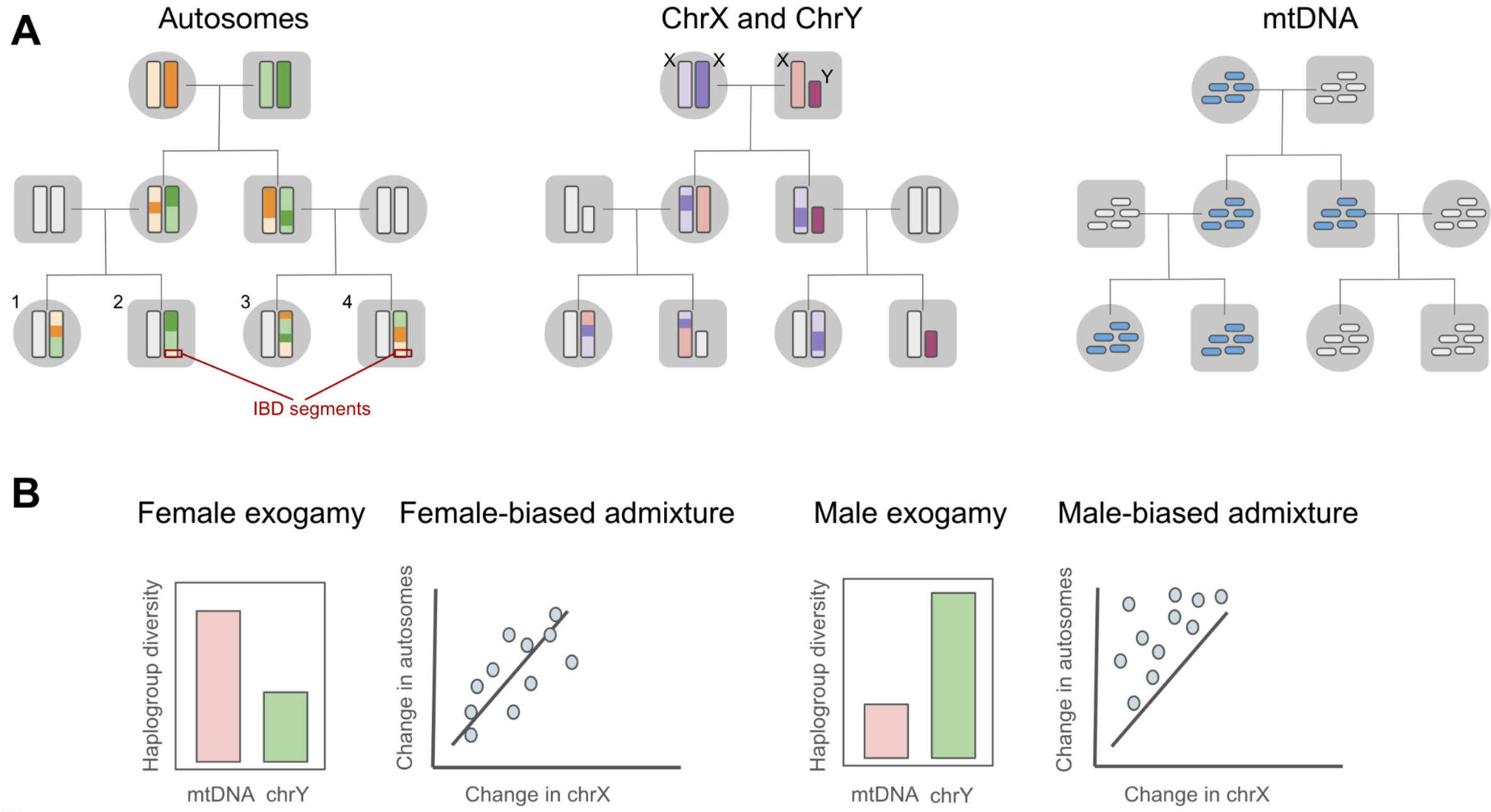

**Figure 2**: Inheritance patterns of genetic markers and genetic signatures of sex-biased mobility. **A)** Three-generation pedigrees showing inheritance of autosomes, chromosome X (chrX), chromosome Y (chrY), and mitochondrial DNA (mtDNA). Squares represent males and circles females. Genetic markers of the focal grandparent pair are coloured, while those of other individuals are shown in grey. Different colours indicate paternal and maternal chromosomal copies; recombination of autosomes and female chrX is illustrated by colour mixing in the next generation. In contrast, mtDNA and chrY are inherited uniparentally without recombination. The autosomal panel represents one of 22 autosomal chromosome pairs and highlights an identical-by-descent (IBD) segment (light orange) shared by individuals 2 and 4, inherited from their grandmother. In the mtDNA panel, multiple copies are shown to reflect the high copy number of mtDNA per cell. **B)** Schematic illustration of expected genetic patterns under female and male exogamy, including diversity in mtDNA and chrY lineages and differential genetic change on autosomes versus chrX under sex-biased admixture.

**Alt Text:** Two-panel figure illustrating inheritance of genetic markers and signatures of sex-biased mobility. Panel A shows three-generation pedigrees for autosomes, sex chromosomes (X and Y), and mitochondrial DNA (mtDNA). Coloured chromosomes trace inheritance from a focal grandparent pair, while grey chromosomes represent other ancestry. Autosomes and female X chromosomes recombine across generations, creating mixed-colour chromosome segments; an identical-by-descent segment inherited from a grandmother is highlighted in two grandchildren. In contrast, Y chromosomes are passed from father to son and mtDNA from mother to all children without recombination. Panel B shows expected genetic patterns under sex-biased mobility: female exogamy produces higher mtDNA than Y-chromosome lineage diversity, whereas male exogamy produces higher Y-chromosome than mtDNA diversity. Sex-biased admixture is illustrated by differing levels of genetic change on autosomes and chromosome X.

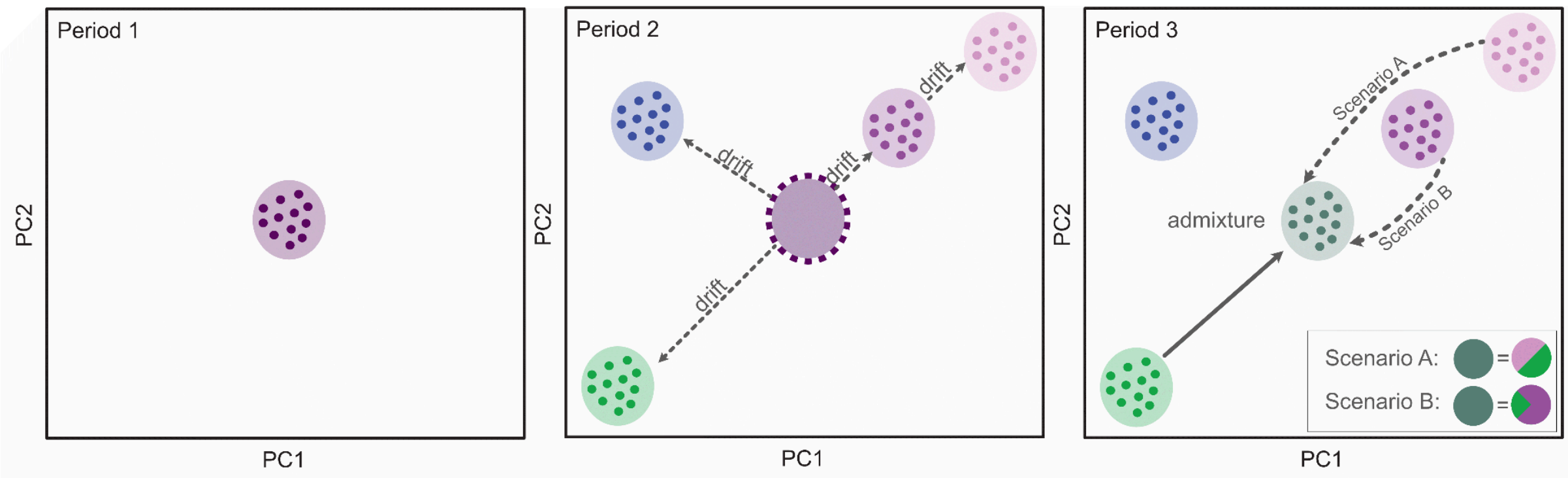

**Figure 3.** A schematic of drift and admixture in principal components analysis space. Here the PC space is the same across time and represents diversity among all temporal populations (i.e. the PC space could be calculated by joining together samples from all time periods). The inset of the right lower panel shows two alternative admixture scenarios that could explain the creation of the lilac population; the size of the circles indicate the relative contributions.

**Alt text:** Three-panel schematic PCA plot showing population relationships through time. In Period 1, a single cluster of dark purple points is centred in the plot. In Period 2, this population has diverged into four spatially separated clusters (blue, green, purple, and pink), connected to the original central cluster by dashed arrows labelled "drift." In Period 3, the four clusters remain distinct. A new teal cluster appears between the green and purple/pink groups, connected by a solid arrow from the green cluster and labelled "admixture." Dashed arrows indicate two possible origins for the teal cluster: Scenario A derives the second ancestry source from the pink cluster, whereas Scenario B derives it from the purple cluster. An inset illustrates the relative ancestry contributions for the two scenarios using coloured circles. The schematic demonstrates equifinality, whereby similar positions in PCA space can arise from different admixture histories and ancestry proportions.

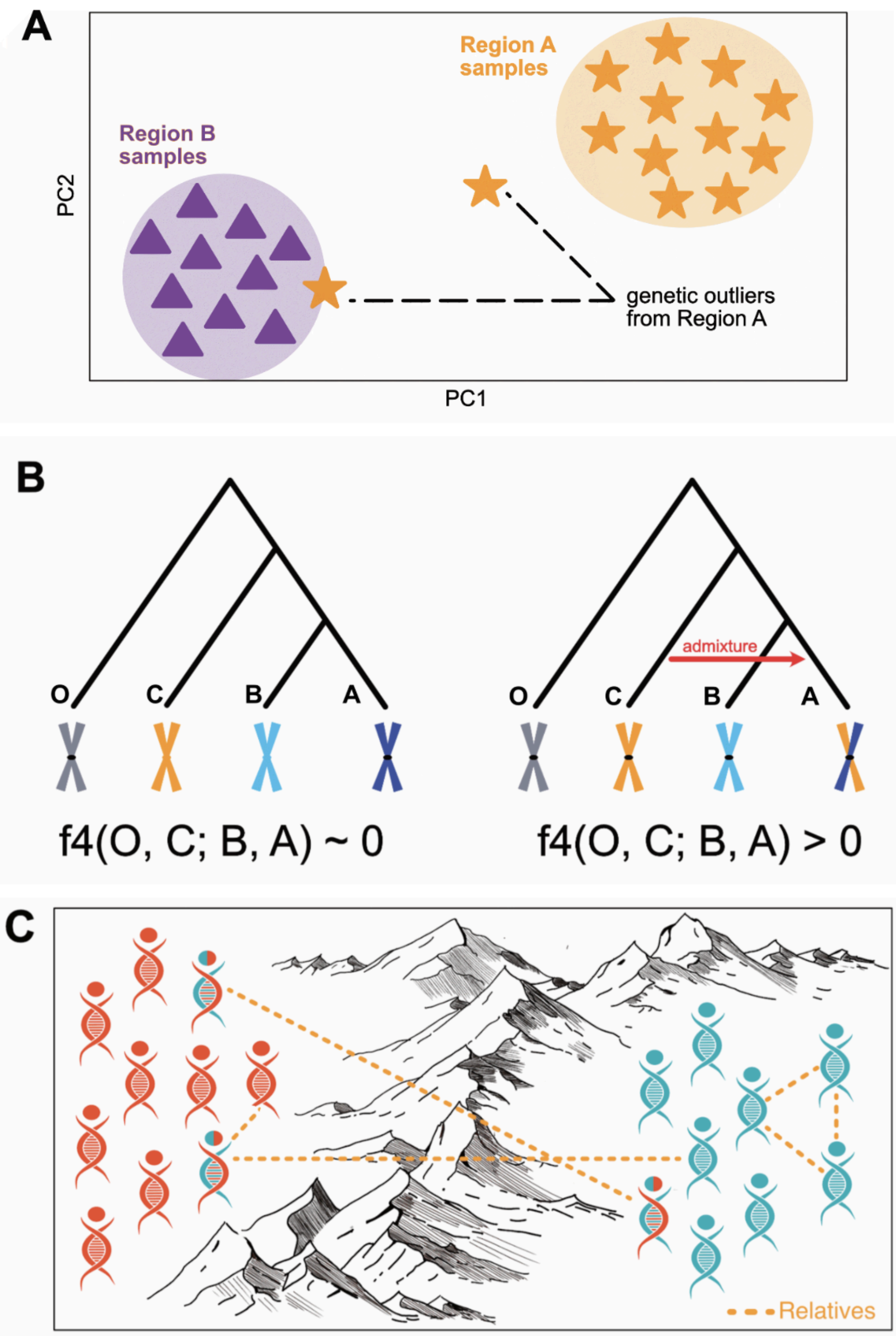

**Figure 4:** Different genetic signatures of mobility. **A)** A schematic example of genetic outliers in PC space. The shapes (X and O) indicate the region an individual was recovered, while the location in the PC space indicates the genetic profile. The two outliers from O could represent a first-generation migrant and an admixed later-generation migrant. **B)** A schematic of the $f_4$ test. The left tree displays a demographic scenario where the common ancestor of A and B have split from that of C, such that C is phylogenetically equidistant to A and B. The outgroup lineage is depicted as O. The right tree shows the same demographic scenario but where C has admixed into A only, which results in higher correlation (affinity) between A-C than between B-C, which is captured by a significantly positive f4 statistic. **C)** A schematic of genetic relatives buried in space. The lines indicate genetic relatedness. Relatives separated by long distances (shown with bolder lines) indicate mobility.

**Alt text:** Three-panel figure illustrating different genetic signatures of mobility. Panel A shows a schematic PCA plot showing points (in yellow and purple) representing the genomes of individuals from two regions (or archaeological sites, in the case of aDNA). Individuals from Region B (purple) are relatively homogenous and all plot in the lower left quadrant of the plot. While many individuals from Region A (yellow), plot in the upper right hand corner of the PCA, two plot elsewhere (including one with the Region B cluster) to represent individuals who are genetic outliers. Panel B shows two simplified population trees used to illustrate the $f_4$ test. In the first tree, each population is presented by a colored chromosome (grey (for the outgroup (O)), yellow (C), light blue (B), dark blue(A)) with no gene flow indicated between groups. f4(O, C; B, A) ~ 0 is shown below, to indicate the $f_4$ statistic would be about zero in this case. The second tree is similar, but a red horizontal arrow labelled “admixture” extends from the branch leading to C toward the branch leading to A. The chromosome icon beneath A is coloured partly yellow and partly dark blue. The equation f4(O, C; B, A) > 0 is shown below to indicate there has been gene flow between C and A. Panel C depicts the same mountainous landscape from Figure 1 with groups of individuals on opposite sides of the mountains. Many individuals on the left are red (8 are red and 2 are half red, half light blue, indicating admixture), and many on the right are light blue (8 are light blue and 1 is half red, half light blue). Dashed orange lines connect genetically related individuals within each population and across the mountain range.

**Supplementary Information**

**Glossary**

admixture: The mixing of genetically distinct populations, detectable through ancestry components from multiple source populations in an individuals/populations genome.

allele: one of two or more alternative versions of a genetic variant at a specific genomic position.

ancestry component: the proportion of genetic variation in an individual or population that is statistically inferred to derive from a specific ancestral source or gene pool. Ancestries usually represent genetic profiles of a particular period and region. Admixed individuals/populations will contain multiple ancestry components at varying levels. These components are model-based constructs and do not necessarily correspond to discrete biological or cultural groups.

ancient DNA (aDNA): refers to DNA retrieved from archaeological, historical, or palaeontological remains, typically highly fragmented and damaged due to post-mortem degradation. The field emerged in 1984 with DNA extraction from dried equid tissue, expanded to human remains by 1989, and achieved whole-genome sequencing of ancient hominins by 2010. Early studies focused on uniparental markers (mtDNA and Y-chromosome), while modern approaches utilize genome-wide data from hundreds of thousands of SNPs to investigate ancestry, admixture, and migration.

ascertainment bias: a trend for better representation of genetic diversity among certain populations when genomic data production targets specific SNPs (as in SNP capture): populations used to identify variable positions will artificially appear more diverse.

autosomes: In humans, chromosomes 1-22, which are generally found in two copies in the genome (except for cases like Trisomy 21, where chr21 is three copies).

ancestry cline: A pattern of genetic similarities changing monotonously over space. This can be created by local mobility and admixture, or also by a migration event that loses strength over time/generations.

coverage: (or depth-of-coverage in this article's context) is the measure of how many times (through how many DNA molecules) each position in the genome is represented in a sequencing dataset. 10x coverage would mean each position is represented by 10 molecules on average, while 0.1x coverage means that roughly 90% of the genome is not observed even once.

cryptic population structure: population structure (the presence of distinct subpopulations within a region) that is not yet identified.

culture-historical theory: the dominant theoretical approach to interpreting archaeological evidence in the late 19th and early 20th century. It arose, in part, to describe and classify the diversity of material culture found across space and time, and rejected the shared, unilinear stages of evolutionary archaeological framework. This approach often essentialized people into ethnic groups defined by the material traces they left behind in the archaeological record (hence equating "pots with people"), inhabiting discrete "cultural zones" on maps. Change was often ascribed to migration, replacement, and diffusion events. This theory became entangled with nationalist histories in Europe in the first half of the 20th century and was replaced by the scientific turn of "New Archaeology" (now commonly known as processual archaeology) in the 1940s-1960s.

diploid: the state of carrying two chromosomes and thus two independent pieces of genetic information (from maternal and paternal origin) at a locus.

downsampling experiments: computational simulations where researchers create subsets of a real genomic dataset (e.g. 0.1x data from an individual with 1x data), perform certain analyses at various genomic coverages, and compare the results from downsampled sets (e.g. 0.1x) with the original observations based on the full data (e.g. 1x). This is a simple way to explore the impact of limited coverage (i.e. missing data) on downstream inference.

f-statistics: a set of statistics ($f_2$, $f_3$, and $f_4$) calculated from genome-wide correlations of allele frequencies between pairs of individuals or populations, indicating genomic similarity or relative similarity, and using in admixture testing and qpAdm modeling.

FST: a measure of genetic differentiation between populations relative to their diversity, also called the Fixation index. It can be measured per variable position.

gene flow: the movement of alleles between gene pools (populations) through migration of individuals or through dispersing gametes (e.g. pollen in plants), resulting in changes in allele frequencies.

gene pool: the total collection of genetic variations present in a population at a given time. In population genetics, it is often modeled abstractly as a collection of allele frequencies, assuming random mating and ignoring spatial or sociocultural structure.

genetic drift: stochastic change in allele frequencies in populations caused by normal biological randomness, e.g. segregation during meiosis, fertilization, chance events impacting reproduction and survival. Drift contrasts with selection, which can cause non-random shifts in allele frequencies.

genetic outlier: an individual who carries a genetic profile different from others in their region and period. This can suggest mobility of that individual or their ancestors, or admixture from a genetically distinct population not well-presented in the regional sample.

genotype: information carried in the DNA. The term is often used to refer to the nucleotide information (A, T, C, G) at a particular chromosomal position of an individual, and can be diploid (e.g. CA) for autosomes and the X chromosome in females, and haploid (e.g. C) for mitochondrial DNA, the Y chromosome, and the X chromosome in males. The term is also used as a verb, referring to decoding the genotype information from DNA sequence data.

haploid: the state of carrying a single chromosome and thus one piece of genetic information (either maternal or paternal origin) at a locus.

haplotype: a chromosome segment (string of DNA) inherited together on the same chromosome across generations. Haplotypes can span short or long genomic regions depending on recombination history.

identical-by-descent (IBD) segments: segments of DNA shared between two individuals that are inherited from a recent common ancestor. For instance, siblings share on average 1/2 of their genome in IBD segments (which can increase or decrease due to meiotic randomness).

imputation: filling in unobserved DNA sequence data using a reference panel of fully sequenced and phased genomes.

isolation-by-distance: the phenomenon that populations of a species found farther apart tend to be genetically more differentiated, owing to the fact that the chances of mating between individuals in distant populations is less likely.

kinship coefficient: an estimate of the probability that randomly chosen loci in two individuals (one from each) are inherited from a recent common ancestor. The expected kinship coefficient for monozygotic twins is 0.5, for first-degree relatives (parent-offspring or siblings) it is 0.25, and for second-degree relatives (e.g. aunt-nephew or half-siblings) is 0.125.

low coverage: when only a small fraction of an individual's genome is represented. There is no strict definition but 0.2x may be considered low.

material culture: the evidence about human life represented in physical objects.

metagenomic data: genomic data that represents multiple species simultaneously, e.g. DNA from mixed microbial species represented in soil samples.

mitochondrial DNA (mtDNA): The DNA of a small cellular organelle that is structurally and evolutionarily distinct from chromosomes found in the nucleus. It is haploid and only inherited through the mother. Its high copy number per cell, relatively high mutation rates, and haploidy (thus ease of analysis) made it the locus of choice in the pre-genomic of ancient DNA.

outgroup: a lineage that is phylogenetically equally distant to all other groups used in an analysis. For instance the chimpanzee is an outgroup to all human populations.

outgroup(f3): a formal test measuring shared genetic drift between two individuals/groups relative to an outgroup. The outgroup(f3) is also the name of the statistic and represents a measure of genetic similarity.

polymorphism (variant): the presence of multiple versions (alleles) at a locus within a population. Polymorphisms are created by mutations.

population structure: heterogeneity in the gene pool, such as two or more groups where individuals mate within groups but not among groups.

pottery chronology: before the advent of radiocarbon dating in the mid-20th century, using material culture artifacts for regional comparisons and creating chronologies was essential in archaeological research, and it remains an important method in archaeological research and interpretation today. Pottery styles (both the form of ceramics themselves, as well as decorative elements) became particularly important for this, due to the pervasiveness and durability of of ceramics in many archaeological contexts and ability to create seriations of ceramic forms, as they usually changed over time such that layers could be dated by the ceramic styles found there.

pseudohaploid: representation of an individual's genotype at a diploid locus with a single allele by randomly sampling a single sequencing read per position. Pseudohaploid genotype data for biallelic SNPs can be thought of as 0s and 1s, representing presence/absence of alternative alleles on either chromosome. Calling pseudohaploid genotypes leads to loss of information, but is a simple solution to the problem of heterogeneous coverage in ancient genomes. It ensures equal representation of all genomes irrespective of their depth-of-coverage.

recombination: the biological mechanism that occurs during gamete production in sexually-reproducing organisms that swaps parts of a diploid individual's maternal and paternal chromosomes. Recombination and Mendelian segregation (random assortment of chromosomes to gametes) ensure that the DNA information passed on to the next generation in gametes is an amalgamation of maternal and paternal DNA information. Thanks to recombination and segregation, every individual carries different demographic histories in their genomic sequence inherited from large numbers of ancestors (e.g. c.1000 ancestors 10 generations ago, assuming no inbreeding).

reference bias: a trend for higher representation of alleles carried on the reference genome that may be caused by alignment algorithms and data filtering procedures, as well as laboratory protocols such as enrichment.

runs of homozygosity (ROH): regions of the genome of an individual that have been inherited from a recent ancestor and are homozygous. If individuals in a population carry an excess of short ROH (relative to other populations), this is indicative of endogamy. If an individual has long ROH, this is indicative of consanguinity.

shotgun sequencing: sequencing DNA fragments randomly without selection or enrichment, as opposed to SNP capture.

significance testing: methods to estimate reproducibility of statistical signals given the variability in the data. In genomic methods described here significance testing largely relies on finding reproducible signals across the genome.

single nucleotide polymorphisms (SNPs): variable positions in the genome where individuals differ by a single nucleotide in a DNA molecule (e.g. from G->A). They are the most common type of genetic variants in population genetics analyses.

SNP capture: DNA sequencing that enriches specific SNP positions within the genome. It involves binding ("capturing") DNA fragments with probes, and is preferred because it lowers sequencing costs. The 1240k and TWIST SNP capture kits have been widely used and constitute the bulk of ancient human individuals genetically analysed yet. The approach has been criticized for creating biased representations of DNA data, as opposed to shotgun sequencing.

statistical phasing: the estimation of the sequence of the two chromosomes (one maternal and one paternal) of an individual using information from correlations of alleles among individuals in a large reference panel.

Strontium ($^{87}Sr/^{86}Sr$) isotope analysis: involves measuring the ratio of two strontium isotopes ($^{87}Sr/^{86}Sr$) that tend to vary among geological regions, which are incorporated into our tissues from food and drinking water. The ratio, measured from bones or teeth, can then be an indicator of where an individual was located during the development of those tissues.

third-degree relatedness: genetic relatedness that includes cousins or great-grandparents and great-grandchildren.

uniparental markers: markers transmitted only from one parent: mitochondria from mothers to all children, and chromosome Y from fathers to sons. These markers trace only a single ancestral line (maternal or paternal) and therefore represent a small fraction of an individual's ancestry.

X chromosome (chrX): The sex chromosome which is diploid in females but haploid in males (in humans and most other mammals).

Y chromosome (chrY): the sex chromosome carried by males only. It is haploid and does not recombine.